# Engineering Economy: A New Paradigm for Escaping the Middle-Income Trap

## — Lessons from Türkiye and South Korea —

**Mustafa Ergen**

*Professor at Istanbul Technical University*

*Correspondence: mustafaergen@itu.edu.tr*

April 2026

## Abstract

This paper introduces the concept of Engineering Economy as a new paradigm for understanding and managing macroeconomic policy in middle-income countries seeking to escape the middle-income trap. Drawing on Türkiye's post-2001 economic trajectory and South Korea's successful transition from a low-income to a high-income economy, the study argues that conventional frameworks—whether the Washington Consensus's market liberalization prescriptions or the institutionalist critique alone—are insufficient. Instead, it proposes treating the economy as a dynamic control system requiring continuous calibration rather than static equilibrium. The paper develops a road-surface metaphor (highway, side-road, off-road) to characterize different global economic regimes and presents eleven interconnected policy pillars spanning venture capital formation, regulatory sandboxes, technology-focused industrial policy, and human capital development. By synthesizing insights from endogenous growth theory (Romer), institutional economics (Acemoğlu), the catching-up literature (Lee), cybernetic systems theory (Wiener), and Schumpeterian creative destruction, the framework reconceptualizes macroeconomic instruments through control-engineering analogies: interest rates as energy gradients, fiscal policy as energy flow, exchange rates as balance motors, and regulation as adaptive suspension. The analysis demonstrates that Türkiye's structural challenge is not merely institutional weakness but a systemic absence of R&D demand from its dominant enterprise structures, creating a vicious cycle that conventional reforms cannot break. Seven specific opportunity windows arising from US–China technological rivalry are identified, and a phased implementation roadmap is proposed.

## 1. Introduction

The global economic landscape has undergone a fundamental transformation since the 2008 financial crisis. The era of hyper-globalization, during which the world trade-to-GDP ratio climbed steadily from approximately 37% in 1986 to 61% in 2008 (World Bank, WDI), has given way to what economists' term 'slowbalization'—a period of stagnating trade openness, rising protectionism, and fragmenting supply chains (Hoekman, 2015). By 2023, the global trade-to-GDP ratio had declined to approximately 57% (World Bank, WDI), while world GDP growth decelerated from a 2000–2010 average of 3.7% to the IMF's 2025 projection of approximately 3.2% (IMF, 2025). The post-pandemic world economy resembles what this paper characterizes as an 'off-road' terrain: unpredictable, uneven, and demanding adaptive rather than formulaic responses.

The structural shift is compounded by the return of industrial policy in advanced economies. The United States' CHIPS and Science Act (2022) and Inflation Reduction Act (2022), the European Union's Green Deal Industrial Plan, and China's 'Made in China 2025' initiative represent a fundamental departure from the free-trade orthodoxy that prevailed during the hyper-globalization era. For middle-income countries, the competitive landscape has shifted: they must now contend not only with low-cost competitors in developing countries but with massively subsidized high-technology sectors in advanced economies.

For middle-income countries such as Türkiye, this structural shift poses an existential challenge. Türkiye experienced its own 'highway' period between 2002 and 2007, averaging 7.2% annual GDP growth, only to enter prolonged turbulence thereafter. Türkiye possesses many prerequisites for successful catch-up: a large domestic market of over 85 million consumers, a strategic geographic location bridging Europe and Asia, NATO membership, a customs union with the EU, and a diversified manufacturing base. Despite these advantages, Türkiye's GDP per capita has stagnated in dollar terms since 2013, and the country has experienced recurring bouts of high inflation, currency depreciation, and institutional uncertainty—suggesting that the binding constraint lies not in factor endowments but in the absence of a coherent, technology-centered development strategy.

The structural vulnerability of Türkiye's highway-era growth is evident in its composition. Between 2002 and 2008, the construction sector expanded at nearly twice the rate of manufacturing, and the current account deficit widened to nearly 6% of GDP by 2006—financed by short-term capital inflows that proved reversible during the 2008 crisis. The share of high-technology products in total exports remained below 3%, compared to Korea's 30%. In contrast, South Korea's highway-era growth was anchored in export-oriented manufacturing that continuously moved up the value chain: from textiles in the 1960s to steel and shipbuilding in the 1970s, semiconductors and automobiles in the 1980s, and electronics and information technology in the 1990s. Each phase built technological capability that carried forward to the next, creating a ratchet effect absent in Türkiye's construction-led model.

The paradox is particularly acute when viewed comparatively. South Korea, with a population roughly comparable to Türkiye's, no natural resources, a similar post-war starting point, and an even more devastating colonial legacy, achieved high-income status by the early 2000s and now hosts globally leading firms in semiconductors (Samsung, SK Hynix), automobiles (Hyundai-Kia), shipbuilding (HD Hyundai), and consumer electronics (LG, Samsung). Türkiye, by contrast, despite two decades of robust growth in the 2000s, has seen its GDP per capita stagnate in dollar terms since 2013 and retains no globally competitive technology firm. Understanding what drove this

divergence—and what policy framework might close it—is the central analytical objective of this paper.

This paper addresses that gap by introducing the concept of Engineering Economy—a paradigm that treats economic management not as a static optimization problem but as a dynamic control system requiring continuous feedback, adaptive regulation, and strategic calibration. Drawing on control engineering principles and cybernetic systems theory (Wiener, 1948; Forrester, 1961), it argues that macroeconomic instruments function analogously to a vehicle's subsystems: interest rates as traction control (managing the economy's grip on productive activity), exchange rates as a balance motor (maintaining external equilibrium), fiscal policy as energy flow (allocating resources across priorities), and regulation as adaptive suspension (absorbing shocks while maintaining directional stability). This mapping enables a systematic analysis of policy failures as control-system malfunctions: overshooting, oscillation, loss of feedback, or structural resonance.

The contribution is threefold. First, it develops a novel conceptual framework integrating institutional economics, endogenous growth theory, catching-up theory, and cybernetic systems analysis into a single operational paradigm. Second, it provides a detailed comparative analysis of Türkiye and South Korea identifying the specific structural mechanisms through which Korea escaped the middle-income trap while Türkiye stalled. Third, it offers actionable policy prescriptions organized as eleven interconnected pillars with seven specific opportunity windows arising from current geopolitical realignments, moving beyond the diagnostic mode that characterizes much of the existing literature.

The paper also contributes methodologically by importing analytical tools from control engineering and systems theory into development economics. The concepts of feedback latency, control bandwidth, resonance, and energy-gradient flow—standard in engineering disciplines—provide novel purchase on persistent puzzles in development economics, such as why well-designed policies produce oscillatory outcomes, why financial crises in middle-income countries are disproportionately severe, and why institutional reforms that succeed in one context fail in another.

These engineering analogies are not cosmetic; they generate testable predictions about policy dynamics that differ from those of conventional equilibrium models.

The paper proceeds as follows. Section 2 reviews the relevant theoretical literature. Section 3 describes the methodology. Section 4 presents a historical analysis through the road-surface metaphor. Section 5 develops the eleven policy pillars with expanded treatment of regulatory engineering, capital accumulation dynamics, and interest rate theory. Section 6 examines the South Korean case. Section 7 analyzes the R&D demand gap. Section 8 identifies seven opportunity windows from the US–China rivalry. Section 9 discusses findings. Section 10 concludes.

## 2. Literature Review: Growth Theory and the Middle-Income Trap

### *2.1. Endogenous Growth and Human Capital*

The neoclassical growth model (Solow, 1956) attributes long-run growth to exogenous technological progress, leaving the most important driver of prosperity unexplained. Romer's (1990) endogenous growth theory resolved this by demonstrating that investment in research and development generates non-rival ideas whose social returns exceed private returns, justifying public intervention in R&D. For middle-income countries, Romer's framework implies that growth cannot be sustained merely by capital accumulation; it requires building the institutional infrastructure for idea generation.

Lucas (1988) complemented this by emphasizing human capital externalities: economies with higher average skill levels grow faster not only because each worker is more productive but because of spillover effects. The brain drain from middle-income countries represents a devastating loss, as it removes the high-skilled workers whose externalities would benefit the domestic economy most. Aghion and Howitt (1992) extended the Romerian framework by modeling growth as creative destruction, predicting that countries closer to the technological frontier benefit more from innovation-oriented policies, while those further away benefit more from imitation—a distinction directly relevant to middle-income countries occupying an intermediate position.

### *2.2. Institutions and the Washington Consensus*

Acemoğlu and Robinson (2012) argue that inclusive political and economic institutions are the fundamental cause of long-run economic development. The 2024 Nobel Prize recognized this contribution as foundational. The Washington Consensus (Williamson, 1990) prescribed fiscal discipline, trade liberalization, privatization, and deregulation as universal prescriptions. Stiglitz (2002) mounted a sustained critique, arguing that premature liberalization without adequate institutional safeguards devastated developing economies. Rodrik (2006) demonstrated that East Asian economies that liberalized gradually outperformed Latin American economies that adopted the full package simultaneously.

North (1990) provided the foundational theory distinguishing formal rules (constitutions, laws, property rights) from informal constraints (norms, conventions, cultural practices). For middle-income countries, the challenge is that formal institutional reform is often necessary but insufficient: informal institutions—business practices, trust networks, attitudes toward risk and innovation—evolve much more slowly and can undermine even well-designed formal reforms. A country may establish an independent central bank (formal institution) while political culture continues to expect monetary policy to serve electoral objectives (informal constraint). This insight reinforces the Engineering Economy’s emphasis on cultural transformation alongside institutional reform, particularly in building the risk tolerance and entrepreneurial orientation that venture ecosystems require.

North (1990) provided the foundational theory linking institutions to economic performance, distinguishing between formal rules and informal constraints. For middle-income countries, the challenge is that formal institutional reform is often necessary but insufficient: informal institutions—business practices, trust networks, attitudes toward risk—evolve much more slowly and can undermine even well-designed formal reforms. This reinforces the Engineering Economy's emphasis on cultural transformation alongside institutional reform.

### *2.3. Catching Up, Creative Destruction, and Structural Transformation*

The middle-income trap concept gained prominence through Gill and Kharas (2007). Eichengreen, Park, and Shin (2014) found that growth slowdowns are most likely at per capita income levels around $10,000–$16,000 (2005 PPP), often triggered by declining total factor productivity. Lee (2013, 2019) offers the most operationally relevant framework: his 'short-cycle technology' hypothesis posits that latecomers should target domains where patent-to-citation time is short, indicating rapid knowledge turnover and weaker incumbent advantages. Lee identifies two additional conditions: (i) large firms with genuine R&D appetite, and (ii) a critical mass of trained personnel in targeted technology areas.

Lee's empirical evidence is compelling. Analyzing patent data across 130 countries and 44 technology classes, he demonstrates that countries which entered short-cycle technology domains (where the average citation lag was under 5 years) achieved significantly higher catch-up rates than those targeting long-cycle domains (citation lags of 10+ years). The mechanism is straightforward: in short-cycle domains, knowledge becomes obsolete rapidly, meaning that latecomers' disadvantage in accumulated knowledge is smaller and their advantage in agility and lower costs is larger. Blockchain technology, certain AI applications, drone systems, and gaming software exemplify short-cycle domains where Turkish firms have already demonstrated competitiveness—precisely because the incumbents' knowledge advantage is narrow and the pace of innovation rewards adaptability over accumulated expertise.

Krugman's (1991) new trade theory provides theoretical justification for national champion firms capable of achieving increasing returns to scale. Schumpeter (1942) identified creative destruction as the engine of capitalist development; modern evolutionary economics (Nelson & Winter, 1982) extended this into firm-level innovation dynamics. Lewis's (1954) dual-economy model remains relevant for understanding the structural transformation challenge: the persistence of a large, low-productivity traditional sector alongside a small modern sector suggests incomplete structural transformation.

### *2.4. Cybernetic Systems Theory and Economic Management*

A distinctive theoretical contribution of this paper is its integration of cybernetic systems theory into development economics. Wiener's (1948) foundational work on cybernetics established the principles of feedback-based control systems; Forrester's (1961) industrial dynamics applied systems thinking to economic and industrial processes. The Engineering Economy draws on these traditions to reconceptualize macroeconomic management as a control-engineering problem rather than an equilibrium-optimization problem. This perspective aligns with Taleb's (2012) concept of antifragility—systems that gain from disorder—and Perez's (2002) analysis of technological revolutions and financial capital, which demonstrates that each major technological revolution follows a predictable pattern of installation, crisis, and deployment that demands different policy responses at each phase.

Perez’s framework is particularly relevant to the current moment. If artificial intelligence represents a new techno-economic paradigm (analogous to the microprocessor revolution of the 1970s–80s or the internet revolution of the 1990s–2000s), the global economy is currently in the installation phase—characterized by speculative investment, institutional experimentation, and intense competition for positioning. History suggests that installation phases end in a financial crisis (the dot-com bust of 2000, the 2008 crisis following the housing bubble), after which the deployment phase begins under more stable institutional conditions. For middle-income countries, the installation phase offers the best window for catch-up: the technology is still fluid, standards have not yet solidified, and the barriers to entry are lower than they will be once dominant designs emerge. Missing this window—as Türkiye missed the semiconductor window in the 1980s and the internet platform window in the 2000s—could lock the country into technological dependence for the duration of the AI paradigm.

Mazzucato (2013) demonstrated that the state has historically been the primary risk-taker in transformative innovation. Chang (2002) showed that today's advanced economies systematically used protectionist and interventionist policies during their own industrialization phases—policies they now discourage developing countries from adopting. The Engineering Economy synthesizes these heterodox insights with mainstream growth theory under the unifying metaphor of adaptive control engineering.

The cybernetic perspective offers three specific analytical advantages over conventional economic frameworks. First, it foregrounds the concept of feedback latency—the time delay between a policy action and its observable effect. In engineering, feedback latency determines system stability: too much delay causes oscillation and overshoot. In economic policy, the lag between monetary tightening and inflation reduction (typically 12–18 months) creates the same dynamic, explaining why central banks frequently overshoot in both directions. The Engineering Economy's prescription of model-based, forward-looking policy explicitly addresses this problem by incorporating predictive feedback rather than relying solely on lagging indicators.

Second, the cybernetic framework introduces the concept of control bandwidth—the range of perturbations a control system can effectively manage. A central bank operating with a single instrument (short-term interest rate) has narrow bandwidth; a government coordinating fiscal, monetary, trade, and industrial policy has wider bandwidth. The Engineering Economy argues that off-road economic environments demand wide-bandwidth control—precisely the multi-instrument coordination that the Washington Consensus's emphasis on institutional separation (independent central banks, minimal fiscal activism) deliberately constrains.

Third, systems theory distinguishes between negative feedback (stabilizing) and positive feedback (amplifying). Financial markets characteristically exhibit positive feedback: rising asset prices attract more buyers, further raising prices until the bubble bursts. The 2008 global financial crisis was fundamentally a positive-feedback phenomenon that the prevailing economic framework—premised on efficient markets and rational expectations—could neither predict nor manage. Taleb's (2012) antifragility concept extends this insight: rather than merely building robust systems that resist shocks, policymakers should design antifragile systems that improve in response to volatility. This is the deepest aspiration of the Engineering Economy.

## 3. Methodology and Analytical Framework

This study employs a qualitative comparative methodology combining historical-institutional analysis with conceptual framework development. The comparative case-study design follows the 'most similar systems' logic (Przeworski & Teune, 1970): Türkiye and South Korea share strikingly similar starting conditions—comparable population sizes, absence of significant natural resources, post-war reconstruction challenges, authoritarian governance during early industrialization—yet diverged dramatically in outcomes, allowing the analysis to isolate the policy variables that drove the divergence.

The empirical base draws on three data streams: (i) macroeconomic time-series data from the World Bank, IMF, and WTO for global trade-to-GDP ratios, GDP growth rates, and trade volumes from 1960 to 2025; (ii) country-level structural data for Türkiye and South Korea; and (iii) qualitative evidence from the policy histories of both countries, drawing on Lee (2013, 2019), Rodrik (2006), and primary documentation.

The analytical framework centers on a road-surface metaphor that maps global economic regimes to driving conditions. This is operationalized through two quantitative indicators: the global trade-to-GDP ratio (measuring openness) and the volatility of year-on-year GDP growth across major economies (measuring predictability). Highway periods are characterized by rising trade-to-GDP ratios and low growth volatility; side-road periods by plateauing openness and rising volatility; off-road periods by declining openness and high volatility. The 2020–2025 period scores definitively as off-road on both indicators.

The metaphor is not merely illustrative; it carries analytical content. On a highway, the driver's skill matters less than the vehicle's raw power—growth comes from factor accumulation, and the policy prescription is to remove friction (liberalize, deregulate, attract capital). On a side-road, driver skill begins to matter—active steering, gear selection, and hazard anticipation determine outcomes. On off-road terrain, the vehicle's design matters as much as the driver's skill—suspension travel, ground clearance, four-wheel-drive capability, and the ability to recover from slips become decisive. The mapping to economic policy is direct: off-road economies need not just better policies (driver skill) but fundamentally different institutional architectures (vehicle design)—wider policy bandwidth, faster feedback loops, more resilient fiscal and monetary buffers, and the capacity to shift between policy regimes as terrain changes.

The control-engineering analogy maps macroeconomic instruments to vehicle subsystems. This mapping draws on Wiener's (1948) feedback control principles and Forrester's (1961) system dynamics methodology, enabling analysis of policy failures as control-system malfunctions: overshooting (excessive stimulus leading to inflation), oscillation (stop-go cycles), loss of feedback (institutional capture preventing signal transmission), or structural resonance (policy frequencies matching economic cycle frequencies, amplifying rather than dampening volatility).

## 4. Historical Analysis: Global Economic Regimes as Road Surfaces

### *4.1. The Post-War Highway (1950–1970)*

The Bretton Woods era established a stable macroeconomic architecture: fixed exchange rates anchored to gold, capital controls insulating domestic policy from external shocks, and gradually expanding trade under GATT. Advanced economies experienced sustained growth averaging 4–5% annually, while the trade-to-GDP ratio began its long ascent from approximately 20% in the early 1960s. This was the quintessential highway: smooth, predictable, and rewarding standardized management.

### *4.2. Oil Shocks and Stagflation: The First Side-Road (1970–1980)*

The collapse of Bretton Woods (1971) and successive oil crises shattered the highway's predictability. Stagflation defied the Phillips Curve orthodoxy and demanded fundamentally new policy tools. The transition to floating exchange rates introduced currency volatility as a permanent feature. For developing countries, this period represented the first major test of policy adaptability—one that many failed, accumulating the external debt that would trigger the 1980s debt crisis.

#### *4.3. The Hyper-Globalization Highway (1980–2008)*

The period from the early 1980s through 2008 constituted a new highway: trade liberalization accelerated, the world trade-to-GDP ratio surged from approximately 37% in 1986 to 61% in 2008, global value chains multiplied, and financial flows expanded dramatically. World GDP growth averaged 3.7% during 2000–2010 (IMF, 2018). Türkiye experienced its golden era (2002–2008), averaging 7.2% annual growth following IMF-supported reforms. Inflation fell from approximately 70% in 2001 to single digits by 2005.

However, Türkiye's highway growth was qualitatively different from South Korea's: driven by capital inflows, construction, and consumption rather than technology-led export growth. Between 2002 and 2008, the construction sector expanded at nearly twice the rate of manufacturing, and the current account deficit widened to nearly 6% of GDP—financed by reversible short-term capital. Korea's highway growth, by contrast, was anchored in export-oriented manufacturing that continuously moved up the value chain.

#### *4.4. Slowbalization: The Second Side-Road (2008–2019)*

The 2008–2009 financial crisis produced the steepest trade collapse in recorded history—world merchandise trade fell by approximately 12–15% in real terms within a year (Baldwin, 2009). Although trade volumes recovered, the trade-to-GDP ratio plateaued around 60%, signaling exhaustion of hyper-globalization's propulsive forces. The 2010s brought mounting geopolitical tensions, rising protectionism, and early signs of supply-chain regionalization.

#### *4.5. The Off-Road Era (2020–Present)*

COVID-19, the Russia-Ukraine conflict, energy crises, and escalating US-China technological competition pushed the global economy definitively off-road. Trade-to-GDP declined to 52.2% in 2020 before partially recovering. The IMF projects world growth of approximately 3.2% for 2025. Industrial policy has returned to respectability, strategic autonomy has replaced interdependence as a policy objective, and technological sovereignty has become a matter of national security.

The off-road metaphor captures the interdependence of economic, geopolitical, and technological disruptions. The US-China trade war is a technological competition with implications for semiconductor supply chains and AI development. The Russia-Ukraine conflict restructured global energy markets. COVID-19 exposed just-in-time supply chain fragility. In an off-road environment, these shocks arrive simultaneously and interact unpredictably, making static optimization frameworks fundamentally inadequate. The Engineering Economy's emphasis on real-time adaptive response is a direct consequence of this multi-dimensional volatility.

The return of industrial policy in advanced economies further complicates the landscape for middle-income countries. The United States' CHIPS and Science Act (2022) committed over $280 billion to domestic semiconductor manufacturing and scientific research. The EU's Green Deal Industrial Plan channels hundreds of billions of euros toward clean technology. China's 'Made in China 2025' initiative targets self-sufficiency in ten strategic sectors. For middle-income countries, this means the competitive landscape has shifted fundamentally: they must now contend not only with low-cost competitors in developing countries but with massively subsidized high-technology sectors in advanced economies. The Washington Consensus's prescription of non-interventionist trade policy, designed for a world where advanced economies practiced what they preached, is manifestly inadequate for a world where the preaching has stopped.

## 5. The Engineering Economy Framework: Eleven Policy Pillars

The Engineering Economy framework synthesizes theoretical insights into an operationally coherent policy architecture. Unlike the Washington Consensus, which prescribed a fixed set of reforms regardless of context, this framework treats each policy instrument as a component in a dynamic control system. The eleven pillars are organized into three clusters: capital transformation (Pillars 1–3), strategic positioning and macroeconomic calibration (Pillars 4–8), and capability building (Pillars 9–11).

### *5.1. From Old Money to New Money: Venture Capital Formation*

The first pillar addresses the transformation of existing wealth into productive risk capital. In middle-income economies, capital is concentrated in traditional sectors—real estate, construction, commodity trading—controlled by holding groups with little incentive to invest in high-risk technology ventures. The Engineering Economy proposes institutional mechanisms that channel legacy capital into venture funds through tax incentives for LP investments, co-investment structures with sovereign wealth funds, and regulatory frameworks that de-risk early-stage investment. Israel's Yozma program (1993) and Singapore's sovereign wealth fund model demonstrate that state-catalyzed venture ecosystems can overcome market failures—information asymmetries, long time horizons, high failure rates—that deter private capital in markets without established track records (Lerner, 2009). Türkiye's challenge is greater because the underlying business culture, rooted in holding-group paternalism and relationship-based banking, is more resistant to the adversarial, failure-tolerant logic of venture investing.

The venture capital ecosystem follows its own positive-feedback dynamics. A small number of successful exits (IPOs or acquisitions of venture-backed firms) generates two critical resources: experienced entrepreneurs who become angel investors or venture partners, and a demonstration effect that attracts new entrants. Silicon Valley's dominance in global venture capital is not the result of superior talent or policy but of a 60-year accumulation of such feedback loops. Middle-income countries cannot replicate this history, but they can accelerate the cycle by importing experienced venture capital managers (through fund-of-funds mandates requiring international GP partnerships), creating artificial exit opportunities (government procurement preferences for venture-backed firms), and reducing the regulatory friction that deters fund formation (simplified GP/LP structures, tax pass-through treatment for carried interest).

### *5.2. Electroshock Technology Partnerships*

Where domestic R&D capacity is insufficient, strategic partnerships with global technology leaders serve as a defibrillator for the innovation system. The model is not passive technology

transfer—the license-and-assemble approach that characterized Turkish holdings' relationship with foreign partners—but active co-development with mandatory knowledge transfer, joint patent ownership, and local hiring requirements. Samsung's trajectory from licensing basic semiconductor process technology from Micron in the 1980s to surpassing its former licensor within two decades demonstrates that partnership can be a launchpad for indigenous capability if institutional incentives align (Lee, 2013).

### *5.3. Regulatory Engineering: The Sandbox as Adaptive Suspension*

Regulation in the Engineering Economy is not a fixed rulebook but an adaptive suspension system that absorbs terrain shocks while maintaining directional stability. This pillar introduces three performance metrics for regulatory systems: speed (time from application to market authorization), trust (measured through adoption rates and complaint indices), and adaptability (frequency of regulatory updates to technological change). The regulatory sandbox—a controlled environment where innovative products can be tested under relaxed requirements—is the primary instrument.

The framework extends the sandbox concept in three ways. First, co-regulation: the state acts as a co-designer alongside industry, defining minimum safety standards while leaving implementation flexibility to firms. Second, controlled data permeability: regulatory frameworks that allow data to flow across sectors and borders under calibrated privacy protections, recognizing that excessive data localization stifles AI development while uncontrolled flows create security risks. Third, global regulatory diplomacy: middle-income countries can attract investment by offering regulatory environments that are simultaneously compliant with EU and US standards but faster and more flexible in execution—becoming 'regulatory bridges' between competing normative blocs (Ergen, 2025).

### *5.4. Strategic Positioning Amid US–China Technological Rivalry*

The intensifying US–China technological competition creates both risks and opportunities for middle-income economies positioned as 'connector states'—countries maintaining productive relationships with both sides of the geopolitical divide (Farrell & Newman, 2019). This paper identifies seven specific opportunity windows for strategically positioned middle-income countries:

First, an AI Act Refugee Sandbox: the EU's comprehensive AI Act creates compliance burdens that may drive AI startups to seek more agile regulatory environments. A middle-income country offering a fast-track AI sandbox aligned with EU safety principles but with shorter approval timelines could attract significant investment. Second, the German Energy Transition Gap: Germany's accelerated deindustrialization in energy-intensive sectors (chemicals, metals, glass) creates opportunities for countries with lower energy costs and adequate industrial infrastructure to absorb displaced production capacity. Third, Gaming and Digital Content Restrictions in China: Beijing's tightening controls on gaming hours, content, and foreign investment in digital entertainment are pushing Chinese studios and global publishers to relocate development capacity to countries with liberal content regimes and available talent pools.

Fourth, a Crypto and Digital Finance Safe Harbor: as the US and EU tighten cryptocurrency regulation, a well-designed digital finance sandbox could attract blockchain infrastructure and fintech innovation. Fifth, Biotech Fast-Track Trials: countries offering streamlined clinical trial approval processes—while maintaining safety standards—can attract pharmaceutical and biotech R&D investment that would otherwise queue for years in FDA or EMA pipelines. Sixth, US Federal Research Funding Cuts: recent reductions in US federal research budgets are creating a diaspora of scientists and research teams seeking alternative institutional homes. Seventh, Apple and Tech Supply Chain China+1 Diversification: as major technology companies implement 'China plus one' strategies, countries with customs union access to European markets, competitive labor costs, and existing automotive/electronics manufacturing ecosystems are natural candidates for alternative production sites.

Each of these seven opportunity windows has a limited time horizon. The AI Act refugee window will close as other jurisdictions develop their own AI sandbox frameworks; the German energy gap will narrow as firms complete their relocations; the gaming content window will shift as Chinese regulations evolve. The Engineering Economy's emphasis on rapid iteration and three-to-five-year experimentation cycles is specifically designed for this kind of time-limited opportunity exploitation. A conventional industrial policy process—multi-year planning, legislative authorization, institutional creation, program design, application review—would consume most of the window before producing any output. The Engineering Economy demands a different operating tempo: executive-level decision authority, pre-authorized funding mechanisms, and regulatory fast-tracks that can deploy resources within months rather than years.

### *5.5. Technology Funds over Financial Funds*

Middle-income countries typically develop financial sectors that intermediate savings into real estate, government bonds, and consumer credit—high-margin, low-risk activities contributing little to productivity growth. The Engineering Economy advocates shifting incentives toward technology-focused funds: venture capital, growth equity, and project-finance vehicles mandated to invest in R&D-intensive firms. Tax policy should discriminate in favor of long-term equity holdings in technology firms over short-term financial speculation. South Korea's Korea Development Bank and Korea Technology Finance Corporation provided targeted financing for firms in government-designated strategic sectors, accepting lower returns and longer time horizons than commercial banks—functioning as patient capital providers where private markets would not venture.

The distinction between financial funds and technology funds is not merely sectoral but temporal. Financial funds optimize for quarterly or annual returns, driving capital toward assets with short payback periods—real estate, consumer credit, commodity trading. Technology funds accept five-to-ten-year return horizons, enabling investment in R&D-intensive firms whose value creation is front-loaded in expenses and back-loaded in revenues. The mismatch between Türkiye's financial system (overwhelmingly oriented toward short-term returns) and the capital needs of technology-driven growth (requiring patient, risk-tolerant investment) is a structural constraint that no amount of R&D tax credits can overcome without fundamental reform of the financial ecosystem's incentive architecture.

### *5.6. Inflation as Feedback Signal: Temperature Gauge of the Economic Engine*

Conventional inflation-targeting treats price stability as an end in itself. The Engineering Economy reconceptualizes inflation as a diagnostic signal within the economic control system—analogous to engine temperature in a vehicle. Persistently high inflation signals structural imbalances (supply-side bottlenecks, currency depreciation pass-through, fiscal dominance) that cannot be resolved by interest rate adjustments alone. The policy response must address root causes: supply-chain resilience, energy independence, food security, and productive capacity expansion.

The Turkish experience between 2021 and 2023 provides an instructive case study. The central bank's rate-cutting cycle—reducing the policy rate from 19% to 8.5% while inflation was accelerating—violated both conventional and Engineering Economy prescriptions. The conventional prescription would have been to raise rates to anchor expectations. The Engineering Economy prescription would have been to coordinate rate adjustments with supply-side interventions targeting the specific sectors driving inflation (energy, food, housing) while maintaining overall monetary credibility. Instead, the uncoordinated rate cuts created a positive feedback loop: lower rates → currency depreciation → imported inflation → erosion of real incomes → further currency pressure. The episode demonstrates that adaptive monetary policy—the Engineering Economy's recommendation—requires not less analytical rigor than inflation targeting but substantially more: a multi-equation model of transmission mechanisms, real-time monitoring of sectoral price dynamics, and institutional credibility sufficient to persuade markets that heterodox policy is strategic rather than arbitrary.

Türkiye's experience with unconventional monetary policy during 2021–2023—deliberate interest rate reductions amid rising inflation—provides a cautionary illustration. Inflation surged above 70% in 2022 and the lira depreciated dramatically. The lesson is not that central bank independence is sacrosanct—the Engineering Economy explicitly challenges rigid institutional orthodoxies—but that departures from orthodoxy require superior analytical frameworks and institutional capacity, not their abandonment. Central bank–Treasury coordination should be model-based and rapidly adaptive, treating inflation as feedback to be interpreted rather than a target to be mechanically pursued.

### *5.7. Interest Rate as Energy Gradient: A Fluid Dynamics Analogy*

The conventional view treats the interest rate as a price—the price of money. The Engineering Economy offers a deeper analogy: the interest rate functions as an energy gradient in a fluid dynamics system. Capital, like fluid, flows from high-pressure zones (surplus economies, risk-averse savers) to low-pressure zones (deficit economies, high-return opportunities). The interest rate differential between countries determines the direction and velocity of this flow. When a middle-income country raises rates, it increases the 'pressure differential' attracting foreign capital—but simultaneously increases the 'resistance' in domestic economic circuits, slowing investment and consumption.

This analogy reveals a structural problem that conventional analysis obscures: the resonance problem. When external capital flows and domestic credit cycles synchronize—as they did in Türkiye during 2003–2008—the system amplifies small perturbations into large oscillations. A modest tightening of global liquidity conditions triggers capital outflows, currency depreciation, imported inflation, and further tightening in a self-reinforcing spiral. The engineering response is to build structural capacitors: foreign exchange reserves, sovereign wealth buffers, and counter-cyclical fiscal mechanisms that absorb oscillations rather than amplifying them. The distinction between financial funds (which amplify pro-cyclicality) and technology funds (which build productive capacity independent of financial cycles) is critical to breaking resonance.

### *5.8. Budget as Energy Flow: Strategic Fiscal Engineering*

In the Engineering Economy, fiscal deficits are engineering instruments whose value depends on what they finance. A deficit funding measurable-return investments in R&D infrastructure, energy transition, or digital connectivity is fundamentally different from one financing consumption subsidies or civil-service expansion. The framework advocates project-based budgeting with explicit return-on-investment metrics, stress-tested against adverse scenarios—treating public expenditure with the same analytical rigor applied to corporate capital allocation.

### *5.9. Exchange Rate as Balance Motor*

Exchange rate management in an off-road environment requires proactive instruments: stress tests, simulation modeling, and real-time monitoring of import–export composition and liquidity balance. The conventional prescription of ‘let the market determine the rate’ assumes a highway environment where market signals are informative. In off-road conditions, market-determined rates can amplify instability. Türkiye's persistent current account deficit—averaging approximately 5% of GDP over two decades—creates structural dependence on foreign capital inflows, making the exchange rate hostage to global risk sentiment. Breaking this pattern requires structural transformation of the trade balance through higher-value-added exports—circling back to the technology and R&D pillars.

### *5.10. Capital Accumulation as Engine Design: The Three-Gear Model*

The Engineering Economy reconceptualizes capital accumulation as an energy conversion problem. An economy’s ‘engine’ converts inputs (savings, foreign investment, natural resources) into outputs (GDP, employment, welfare) with varying efficiency. The Incremental Capital-Output Ratio (ICOR) measures this conversion efficiency: a lower ICOR indicates that each unit of investment generates more output. Middle-income countries typically suffer from rising ICOR as diminishing returns set in—the classic growth slowdown.

Cross-country evidence supports this framework. South Korea’s ICOR averaged approximately 3.0 during its high-growth decades (1965–1995), meaning each dollar of investment generated roughly 33 cents of additional GDP. Türkiye's ICOR has deteriorated from approximately 4.0 in the 2000s to over 7.0 in the 2010s, indicating that investment efficiency has halved—each dollar of investment now generates less than 15 cents of additional output. This deterioration reflects the three-gear misalignment: financial capital (Gear 1) continues to flow—largely into construction and consumer credit—but technological capital (Gear 2) remains static and trust capital (Gear 3) has eroded through institutional uncertainty. The result is an engine that consumes fuel (savings and foreign capital) at an accelerating rate while producing diminishing output—the economic equivalent of a transmission slipping between gears.

The framework introduces a three-gear transmission model for capital accumulation. The first gear is financial capital: the volume and cost of investable funds. The second gear is technological capital: the knowledge, patents, and process capabilities that determine how productively financial capital is deployed. The third gear is trust capital: the institutional credibility, rule-of-law quality, and regulatory predictability that determine whether investors commit to long-term, illiquid investments. Effective capital accumulation requires all three gears to mesh: financial capital without technological capability produces asset bubbles; technological capability without trust capital produces brain drain; trust capital without financial depth produces stability without growth. Risk capital functions as the ignition system—the initial spark that starts the engine—while the multiplier effect represents torque: the amplification of initial investment through supply chains, employment, and demand creation. Transparency mechanisms serve as the cooling system, preventing the engine from overheating through corruption, rent-seeking, or speculative excess.

### *5.11. Human Capital in the AI Age*

Human capital policy faces a particular challenge in the AI age. The traditional model of mass higher education producing generalist graduates for administrative employment is obsolete. The Engineering Economy requires an education system that produces risk-tolerant, experiment-oriented individuals from primary school onward, combined with university systems that are autonomous, performance-funded, and deeply integrated with industry. The state's role shifts from direct control of curricula to indirect but powerful steering through R&D funding, student scholarships, and performance-based institutional funding. Korea's approach—investing heavily in technical education from the 1960s, establishing specialized engineering universities, sending thousands abroad for advanced training in targeted fields, then channeling returnees into chaebol R&D departments—created a feedback loop between education investment and industrial capability. Türkiye's contrasting experience, where foreign-trained academics remain in university positions disconnected from industry, pursuing research agendas inherited from doctoral advisors in advanced economies, illustrates the cost of severing this feedback loop.

The education system redesign must begin at the primary level. The dominant pedagogical model in middle-income countries—rote memorization, risk avoidance, single-correct-answer assessment—produces graduates who are institutionally conditioned against the experimental, failure-tolerant mindset that entrepreneurship and innovation require. Finland's education reforms demonstrate that systemic change is possible within a single generation: by shifting to inquiry-based learning, reducing standardized testing, granting teachers professional autonomy, and valuing creative problem-solving over factual recall, Finland produced PISA scores that rival East Asian systems while maintaining the individual initiative and creativity that those systems often suppress. The Engineering Economy requires a similar transformation in Türkiye, recognizing that the human capital pipeline—from primary school through university to the labor market—is a single integrated system whose output quality is determined by its weakest link. However, even the best-designed pipeline requires merit-based institutional authority to function at full capacity. When control processes block meritocratic advancement, the system operates at basal metabolic rate—surviving but not advancing (Ergen, 2026). Authority can be delegated; institutional legitimacy must be built through structures that reward competence and accountability.

## 6. South Korea: A Successful Engineering Economy in Practice

South Korea's transformation from a war-devastated, aid-dependent economy in the 1950s to the world's twelfth-largest economy represents the most compelling case of middle-income trap escape in modern history. Korea is the only country to have transitioned from receiving OECD Development Assistance Committee (DAC) aid to providing it (Lee, 2013). Understanding how Korea achieved this requires examining the Engineering Economy principles it applied—often avant la lettre—at each stage of development.

Korea's approach to financial liberalization exemplifies adaptive engineering. Rather than following the Washington Consensus prescription of rapid financial deregulation, the Korean government-maintained state control over banks for approximately two decades, using directed credit at below-market interest rates to channel domestic savings into manufacturing capacity expansion. The domestic savings rate rose from approximately 3% of GDP in the early 1960s to roughly 35.8% by 1989 (Bank of Korea)—not despite financial repression but partly because of the income growth that directed credit made possible. Banks were privatized only after the industrial base was sufficiently mature to generate market-rate returns.

Trade liberalization followed a similarly engineered sequence. Korea imposed very high tariffs on consumer goods (protecting domestic firms' monopoly rents, which were reinvested in capacity expansion) while maintaining very low tariffs on capital goods that domestic industry needed to import. Crucially, protection was conditional on export performance: firms failing to meet export targets lost preferential treatment. This linkage between protection and discipline—absent in most import-substitution regimes—ensured that domestic firms remained exposed to international competitive pressures even while enjoying domestic market protection.

The chaebol system—Korea's large, diversified conglomerates—functioned as the 'big firms with R&D appetite' that Lee identifies as essential for catching up. Unlike Turkish holding companies, which sourced technology from foreign joint-venture partners and therefore had minimal incentive to develop indigenous R&D capability, Korean chaebols were pushed by government policy to develop their own technological capabilities. Samsung's semiconductor division began by licensing basic process technology from Micron Technology in the 1980s, then systematically invested in indigenous R&D until it surpassed its former licensor within two decades.

Korea's human capital strategy was equally engineering-minded. The government invested heavily in technical education from the 1960s onward, establishing specialized engineering universities and sending thousands of students abroad for advanced training in targeted fields. Critically, returnees were channeled into chaebol R&D departments and government research institutes, creating a feedback loop between education investment and industrial capability. The temporal dimension deserves emphasis: the entire catch-up process unfolded over approximately four decades, with each decade building on the previous one's achievements. Financial liberalization came last—a direct violation of orthodox advice that proved economically rational in retrospect.

## 7. The R&D Demand Gap: Türkiye's Structural Challenge

Türkiye's failure to replicate Korea's trajectory cannot be attributed to a single cause—institutional weakness, premature liberalization, or insufficient R&D spending. Rather, the core pathology is a systemic absence of R&D demand from the dominant enterprise structures, creating a vicious cycle that conventional reforms cannot break.

Türkiye's enterprise landscape is dominated by two types of actors: large holding companies and SMEs. Holdings historically sourced their technology through joint ventures with foreign multinationals—commercially rational but eliminating the incentive to develop indigenous R&D capability. The holding structure itself is not the problem; Korea's chaebols were structurally similar. The critical difference is that Korean chaebols were directed toward R&D and export competition, while Turkish holdings were directed toward domestic market protection and import intermediation. R&D appetite simply never formed.

The difference in holding/chaebol behavior can be traced to a single policy variable: the conditionality of state support. Korean chaebols received subsidized credit, tariff protection, and preferential market access—but these benefits were explicitly conditional on meeting export targets and investing in R&D. Firms that failed to meet conditions were penalized: credit was withdrawn, protection was reduced, and in extreme cases, the firm was allowed to fail. Turkish holdings, by contrast, received similar forms of state support (subsidized credit through state banks, tariff protection through the customs union's external tariff, preferential government contracts) without equivalent performance conditions. The result was a business culture oriented toward maintaining access to state resources rather than building competitive capability—rent-seeking rather than value creation. The Engineering Economy's insistence on conditionality-based support is the policy lever that transforms holding companies from obstacles to engines of technological catch-up.

SMEs, constituting 99.7% of all enterprises and accounting for 70.5% of employment (TurkStat, 2023), operate predominantly in low-value-added sectors—retail trade, basic manufacturing, transportation. Their limited scale and margins leave no room for R&D investment. The result is an economy where neither the large firms nor the small ones generate meaningful demand for university-produced research or domestically developed technology.

This demand gap cascades through the innovation system. Universities, finding no industrial customers for applied research, retreat into basic science or follow research agendas of advanced-economy institutions where their faculty were trained. Türkiye now hosts over 200 universities—among the highest counts in Europe—yet R&D expenditure remains below 1.5% of GDP, compared to South Korea's 4.9%. Scientific publication output ranks in the global top 20 in quantity, but citation impact metrics fall significantly below the OECD average. The disjunction between volume and impact reflects the disconnect between research activity and industrial relevance. The university, starved of industrial relevance, degrades into a credential mill; teaching loads increase, research budgets shrink, and the institution converges toward a polytechnic.

The consequence is a triple hemorrhage of human capital. First, the most talented graduates leave for advanced-economy firms and research institutions that offer both better compensation and more intellectually stimulating work—the classic brain drain. Second, those who remain in Turkish universities are incentivized by promotion criteria that reward international publication over industrial collaboration, reinforcing the disconnect. Third, the rare Turkish firms that do invest in R&D find they must compete for talent with global technology companies offering remote-work arrangements at developed-economy salaries—a phenomenon accelerated by the post-COVID normalization of distributed work. Breaking this triple hemorrhage requires simultaneously creating domestic demand for high-skilled researchers (through R&D-hungry firms), reforming university incentive structures (through industry-collaboration metrics in funding formulas), and offering competitive conditions for returnees (through targeted salary supplements and research infrastructure).

Breaking this vicious cycle requires simultaneous intervention at multiple points. Increasing R&D subsidies alone fails because the holding structure cannot absorb them; expanding research budgets alone fails because the output goes to developed-country research agendas; liberalizing markets alone fails because foreign monopolies eliminate domestic competitors before they mature. Lee’s (2019) concept of ‘detours’ is directly applicable: rather than competing in long-cycle technologies dominated by incumbents with deep patent portfolios, Türkiye should identify short-cycle technological domains—AI applications, drone technology, digital gaming, cybersecurity, certain biotechnology niches—where rapid knowledge turnover weakens incumbent advantages. Türkiye's existing strengths in unmanned aerial vehicles and digital gaming demonstrate that pockets of competitive advantage can emerge even in an otherwise challenging innovation landscape.

The three-to-five-year experimentation cycle proposed in this framework is deliberately short. Rather than committing to decade-long industrial strategies that risk locking resources into declining sectors, the Engineering Economy advocates rapid iteration: identify a short-cycle technology opportunity, invest in building the human-capital base and firm-level capability, evaluate results after three to five years, and either scale the success or pivot to a new domain. This iterative approach mirrors the lean-startup methodology at the firm level, applied at the scale of national industrial policy. It also aligns with Perez’s (2002) observation that technological revolutions create windows of opportunity that open and close within relatively short periods—latecomers who miss the installation phase of a new paradigm cannot easily catch up during the deployment phase when incumbents have consolidated their positions.

## 8. Discussion

The Engineering Economy framework contributes to the growth literature in several ways. First, it bridges the gap between institutional economics (which identifies what is needed but not how to build it) and industrial policy (which identifies what to do but not in what sequence). By treating the economy as a control system, the framework makes sequencing endogenous to the analysis: policy adjusts in the order that the system's feedback signals indicate, not according to a predetermined checklist.

Second, the framework reconciles the apparently contradictory lessons of the Washington Consensus failure and the East Asian success. Both Rodrik (2006) and Stiglitz (2002) criticized the Consensus's one-size-fits-all approach, but neither offered a unified alternative framework. Lee (2013) provided the technology-policy component; Acemoğlu provided the institutional component; Wiener (1948) and Forrester (1961) provided the cybernetic foundations. The Engineering Economy integrates all three under a single operational metaphor that policymakers can use to diagnose their economy's current condition and determine the appropriate response.

Third, the framework addresses a problem the existing literature largely ignores: what to do when the domestic enterprise structure lacks R&D demand. Most innovation policy assumes firms will invest in R&D if the incentives are right. In holding-dominated middle-income economies, this assumption fails because business models are built around imported technology. The Engineering Economy's response—creating R&D-hungry large firms through targeted support, combined with short-cycle technology targeting à la Lee—is a necessary complement to conventional innovation policy.

Fourth, the cybernetic and fluid-dynamics analogies introduced in Sections 5.7 and 5.10 offer analytical purchase that purely economic metaphors lack. The resonance problem—where external capital flows and domestic credit cycles synchronize to amplify perturbations into crises—is well understood in engineering but poorly captured by equilibrium-based economic models. The three-gear model of capital accumulation (financial capital × technological capital × trust capital) formalizes the intuition that growth requires not merely more investment but the right combination of complementary assets.

It is worth situating the Engineering Economy within heterodox development thinking. Mazzucato's (2013) 'entrepreneurial state' thesis argues the state has historically been the primary risk-taker in transformative innovation. Chang's (2002) analysis of how rich countries actually developed challenges the free-market narrative. Perez's (2002) techno-economic paradigm shifts identifies periodic transformations when new technologies restructure economies. Taleb's (2012) antifragility—systems gaining from disorder—resonates with the Engineering Economy's emphasis on adaptive capacity over static optimization. The Engineering Economy shares elements with each but differs in its systematic integration of macroeconomic management, industrial policy, and innovation strategy under a single control-system paradigm.

The political economy dimension deserves attention. The framework requires a developmental state capable of making long-term bets on technology sectors. Korea achieved this under authoritarian governance during its catch-up phase—a context Türkiye should not replicate. How democratic middle-income countries can build strategic economic engineering capacity while maintaining political accountability is among the most important open questions in development economics.

Taiwan's experience offers a partial answer. Taiwan achieved its technological transformation (from textiles to semiconductors) under authoritarian rule but sustained and deepened it after democratization in the late 1980s. The key was the creation of technocratic institutions—notably ITRI (Industrial Technology Research Institute) and TSMC's founding with government backing—whose legitimacy derived from demonstrated competence rather than political patronage. These institutions survived democratization because they had built constituencies of beneficiaries (successful technology firms, skilled workers, global partners) who defended them against political interference. The lesson for Türkiye is that Engineering Economy institutions must be designed from the outset to generate their own political constituencies—through visible, measurable successes—rather than depending on the continued commitment of any particular government.

Limitations should be acknowledged. The framework is developed through qualitative analysis of two cases; econometric validation across a broader sample would strengthen the claims. The control-engineering metaphor risks implying economies can be 'engineered' with physical-system precision—an implication this paper rejects. Economic systems involve human agency, political constraints, and irreducible uncertainty. Future research should explore quantitative calibration of the road-surface taxonomy, agent-based modeling of the R&D demand gap, and applicability to other middle-income countries.

Despite these limitations, the framework illuminates real-world policy dynamics that conventional approaches miss. Türkiye's defense industry—particularly Baykar's Bayraktar TB2 unmanned aerial vehicle, which has been exported to over 30 countries and deployed in conflicts from Libya to Ukraine—demonstrates that short-cycle technology strategies can succeed even in a challenging institutional environment. The TB2's development followed a pattern consistent with the Engineering Economy's prescriptions: visionary entrepreneurs, government demand as initial customer, progressive technological advancement from imported components to indigenous systems, and rapid iteration driven by operational feedback from conflict zones. Similarly, Turkish digital gaming companies (Peak Games, acquired by Zynga for $1.8 billion; Dream Games, most recently valued at ~$5 billion (2025)) represent short-cycle technology successes in a domain where established intellectual property barriers are minimal. These cases suggest that the elements of an Engineering Economy are embryonically present in Türkiye; the challenge is systemic scaling.

Fifth, the Engineering Economy's control-theoretic foundation reveals a striking isomorphism with the emerging paradigm of agentic AI. Agentic AI systems operate through a perception–reasoning–action–learning loop: they observe their environment, form a model of the current state, select an action, execute it, observe the outcome, and update their model accordingly. This is precisely the feedback-control cycle that the Engineering Economy prescribes for macroeconomic management: monitor economic indicators (perception), diagnose structural conditions using the road-surface taxonomy (reasoning), deploy policy instruments calibrated to the diagnosed terrain (action), and iteratively adjust based on observed outcomes (learning). The analogy is not merely formal. As AI systems increasingly mediate economic activity—algorithmic trading, autonomous supply-chain optimization, AI-driven credit scoring, automated regulatory compliance—the economy itself is becoming an agentic system composed of interacting autonomous agents. Managing such an economy with static, rule-based policy frameworks designed for a world of human-speed decision-making is analogous to controlling a Formula 1 car with a horse-drawn carriage's steering mechanism. The Engineering Economy's insistence on adaptive, feedback-driven governance anticipates a future in which economic policymaking must operate at machine speed, using real-time data streams and simulation-based scenario analysis rather than quarterly statistical releases and annual budget cycles. In this sense, the Engineering Economy is not only a framework for escaping the middle-income trap but a prototype for the governance architecture that all economies—including advanced ones—will eventually require as agentic AI transforms the speed, complexity, and interconnectedness of economic activity.

## 9. Conclusion

Türkiye's escape from the middle-income trap requires not merely producing more but fundamentally reconceiving how and why the economy produces. At the center of this reconception stands the Engineering Economy paradigm: an approach that treats macroeconomic management as adaptive control engineering rather than static optimization. The era of smooth macroeconomic highways—where standardized policies drawn from textbooks or international consensus documents could reliably deliver growth—has ended. Today’s global economy is an off-road terrain: volatile, fragmented, geopolitically charged, and technologically disruptive.

Success in this environment requires not speed or brute force but the right vehicle design, the right calibration, and rapid reflexes. At the macro level, Türkiye should target Korea’s industrial discipline; at the micro level, Silicon Valley’s entrepreneurial dynamism. The two are complementary, not contradictory. The eleven policy pillars presented in this paper—spanning venture capital formation, technology partnerships, regulatory sandboxes, strategic geopolitical positioning, technology-focused finance, adaptive macroeconomic management, and human capital transformation—constitute an integrated system, not a menu of independent reforms. Their power lies in their interaction.

The sequencing matters as much as the content. The Engineering Economy framework suggests a phased approach: first, stabilize the macroeconomic environment to create the predictability long-term investment requires (Pillars 6–9); second, create institutional infrastructure for venture capital and technology partnerships (Pillars 1–3); third, identify and invest in short-cycle technology domains with comparative advantage potential (Pillars 4–5); fourth, reform the education and research system to generate the human capital these domains require (Pillars 10–11). Each phase builds preconditions for the next, evaluated through measurable indicators: patent filings in targeted domains, venture capital deal volume, high-technology export share, and R&D expenditure as a percentage of GDP.

Türkiye's advantage lies not in being larger than its rivals but in the capacity to adapt faster—to build the institutional agility, technological ambition, and entrepreneurial culture the Engineering Economy demands. The elements of success are already visible in Türkiye's defense industry, digital gaming sector, and e-commerce platforms. The challenge is to scale these isolated achievements into a systemic transformation. This requires subjective but visionary decisions, tolerance for error, and the willingness to operate the system iteratively until it works—engineering the economy, not merely administering it.

The institutional architecture required for the Engineering Economy merits specific attention. A National Technology Strategy Board—modeled on Korea’s Ministry of Science and ICT or Israel’s Innovation Authority—should coordinate the eleven pillars, reporting directly to the head of government rather than embedded within an existing ministry. This board would maintain a continuously updated technology radar identifying short-cycle opportunity windows, coordinate R&D funding allocation across universities and firms, manage regulatory sandbox programs, and monitor progress against quantitative benchmarks. Its staff should include practicing engineers and technologists alongside economists, reflecting the paradigm’s insistence that economic governance is an engineering discipline.

A complementary institution—a Technology Venture Fund of Funds—should serve as the financial engine. Seeded with public capital and structured to attract private institutional investors (pension funds, insurance companies, corporate treasury departments), this fund of funds would invest in specialized venture capital funds focused on the technology domains identified by the Strategy Board. The fund-of-funds structure insulates public capital from individual venture decisions while ensuring that the overall portfolio aligns with national strategic priorities. International experience—from Israel’s Yozma to Singapore’s Temasek to Korea’s KDB—demonstrates that such institutions can catalyze private venture ecosystems when designed with appropriate governance, sunset clauses, and performance accountability.

The Engineering Economy is a call for intellectual ambition in economic governance. Just as engineering bridges must be designed for specific loads, terrains, and weather conditions—not for idealized textbook physics—economic policy must be designed for the actual conditions facing a specific country at a specific moment. Korea understood this; its economic miracle was the product not of adherence to any universal doctrine but of relentless, pragmatic adaptation. Türkiye's path to high-income status requires the same pragmatic ambition, the same willingness to learn from failure, and the same commitment to treating economic management as an engineering discipline rather than an ideological one.

The stakes are high. The window of opportunity for middle-income countries to achieve technology-led catch-up is narrowing as advanced economies reindustrialize behind subsidy walls, as AI accelerates the productivity gap between frontier and laggard economies, and as geopolitical fragmentation reduces the scope for neutral free-riding on open global markets. Countries that fail to build indigenous technological capability within the next decade may find themselves permanently relegated to the role of commodity exporters and assembly platforms—dependent on foreign technology, vulnerable to supply-chain disruptions, and unable to generate the high-wage employment that sustains social stability. The Engineering Economy is not merely an academic framework; it is an urgent call to action for policymakers who understand that the terrain has changed and that the old vehicle will not get them where they need to go. This is, in essence, the "wolf reflex": not fighting winter but adapting to it—sensing danger early, preparing quietly, and deploying energy at the right moment (Ergen, 2026). Scaling this reflex, however, requires strong leadership converging with cadres equipped with both technical and moral authority.

Korea's R&D expenditure trajectory illustrates antifragility in action: it nearly doubled from 2.5% to 4.2% of GDP between 1997 and 2007, as post-crisis restructuring redirected resources from unproductive chaebol activities to technology investment. The government made industrial policy smarter rather than abandoning it: performance monitoring became rigorous, international benchmarking was institutionalized, and public support criteria shifted from political connections to measurable innovation outputs. This institutional learning under pressure distinguishes successful Engineering Economies from those that merely survive crises without improving.

# MÜHENDİSLİK EKONOMİSİ

## Yeni Tür(k) Kalkınma Modeli ile Orta Gelir Tuzağından Çıkmak

**Mustafa Ergen**

## Özet

Dünya ekonomisini bir döngüsel yolculuk olarak görmeliyiz. Teknoloji, ekonominin hem yönünü hem de zeminini değiştiriyor. Bir dönemde yollar pürüzsüzken, diğerinde çamurlu ve çetin hale gelebiliyor. Bu minvalde dünya ekonomisi uzun süre otobandaydı: Sermaye serbestçe dolaşıyor, teknoloji üretkenliği artırıyor ve küresel ticaret büyümeyi besliyordu. 2000'ler boyunca ticaret/GSYH oranı %60'lara ulaşırken, dünya ekonomisi reel olarak ortalama %3,7 civarında büyüyordu.

Ancak artık yol değişti. Jeopolitik gerilimler, enerji dönüşümü, Covid-19 sonrası tedarik zinciri kırılmaları ve yapay zekâ devrimiyle birlikte küresel ekonomi off-road bir zemine girdi. Ticaret/GSYH oranları düşüyor, büyüme hızları yavaşlıyor ve büyük krizler, otoban düzlüğünün sürdürülemez olduğunu gösteriyor.

Metaforla anlatırsak:

- 140 km/s hız sınırı olan otobanda 100 km/s ile gitmek başarı sayılır; zemin buna uygundur.
- Aynı araçla tali yola son gaz girip toza boğulmak ise başarısızlık sayılabilir. Bugün buradayız.
- Off-road'da ise otoban aracı ile ilerleyemeyiz. Yeni araç, yeni sürüş getirmeliyiz.

Burada amaç zemine uyum sağlamak, çekiş gücünü korumak ve yön duygusunu kaybetmemektir. Ülkemizdeki otoban mantığıyla tasarlanmış ekonomik politikalar, artık ne küresel zemine uyum sağlayabiliyor ne de verimlilik ve teknoloji sıçraması yaratabiliyor.

İşte bu nedenle Türkiye'nin ihtiyacı Mühendislik Ekonomisidir. Bu yaklaşımı kısaca şöyle tanımlıyoruz: Mühendislik Ekonomisi, ekonomiyi statik bir denge problemi değil, sürekli ölçülen, modellenen ve geri beslemelerle ayarlanan dinamik bir sistem olarak ele alan; politika araçlarını (faiz, kur, bütçe, regülasyon, sermaye yönlendirmesi) birbirinden bağımsız kaldıraçlar değil, birlikte çalışan kontrol bileşenleri olarak tasarlayan ve büyümenin motorunu finansman yerine teknoloji-yetenek-risk sermayesi üçgeni üzerine kuran bir ekonomik akıldır.

Bu yaklaşım, orta gelir tuzağını değişen küresel zeminde teknik bir adaptasyon ve tasarım problemi olarak ele alır; zemine göre vites değiştirir, şartlara göre strateji belirler ve sistemi sürekli hızlıca yeniden ayarlar.

## 2020 ile Dünya Ekonomisi Off-road'a Girdi!

2020 yılında Covid-19 salgını, dünya ekonomisini 2008 krizinden bu yana sürmekte olan tali yoldan çıkarıp tamamen off-road bir araziye attı. Küresel tedarik zincirleri çöktü, üretim giderek yerelleşmeye başladı. Jeopolitik kırılmalar — Rusya-Ukrayna, Çin-Tayvan gerilimleri ve İran savaşı ekonomiye yeni risk katmanları ekledi. Aynı zamanda yapay zekâ, enerji dönüşümü ve veri ekonomisi üretim yapısını yeniden tanımladı.

ABD'nin ilk 10 şirketinin büyük çoğunluğu artık teknoloji odaklı firmalardan oluşuyor; örneğin NVIDIA'nın piyasa değeri pek çok ülkenin GSYH'sini aşmış durumda. Büyük şirketler, yapay zekâ nedeniyle on binlerce çalışanını işten çıkarmaya başladı. Küresel ticaretin GSYH içindeki payı ise %60 bandının zirvesinden %57-58 bandına gerileyerek, küresel sistemin artık tek yönlü olmadığını, çok kutuplu, parçalı ve deneysel bir yapıya dönüştüğünü gösterdi.

2008'e kadar küresel ticaretin GSYH'ye oranı hızla yükseldi; 1986'da yaklaşık %37 olan bu oran, 2008'e gelindiğinde %61 seviyelerine ulaştı. 2022'de küresel ticaret toplamı GSYH'nin yaklaşık %63'üne denk gelmiş olsa da, 2023–2024 verileri düşüş eğilimini gösteriyor; örneğin 2023'te yaklaşık %57 olarak kaydedildi (World Bank, WDI). Bu durum, 2000'ler boyunca artan ticaret/GSYH oranının küresel entegrasyonu ve otoban düzeyindeki ekonomiyi işaret ettiğini ortaya koyuyor.

Dünya ekonomisinin büyüme hızı da benzer bir trend izliyor. 2000–2010 döneminde reel ortalama büyüme yaklaşık %3,7 iken, 2010 sonrası düşüş ve dalgalanmalar gözlemleniyor. 2022'de büyüme %3,24 seviyesinde gerçekleşti ve IMF'ye göre 2025'te dünya ekonomisinin yaklaşık %3,2 civarında büyümesi bekleniyor. Bu yavaşlama, klasik otoban vizyonunun sürdürülemez olduğunu ve küresel sistemin artık daha değişken, yavaş zeminlerde ilerlediğini gösteriyor.

2008–2009 finansal krizi sırasında dünya ticareti yaklaşık %10 oranında çökerken, GSYH de daralmaya girdi. Bu çöküş, küresel entegre otoban sisteminin kırılganlığını ve sınırların, korumacı politikaların ve tedarik zinciri zayıflıklarının etkisini açıkça ortaya koydu.

Özetle, 2000'lerin ortasında ticaret/GSYH oranları hızla yükseliyor ve büyüme istikrarlı görünüyordu; 2010 sonrası ise bu oran zirveden aşağı hareket etmeye başladı, büyüme yavaşladı ve krizlerle kırılganlık arttı. 2008 krizi sistemi tali yola düşürdü, 2020 Covid şoku ise doğrudan off-road'a geçirdi. Türkiye hâlâ dış ticaretle entegre, ama o entegrasyon kırılganlaşıyor. Ticaret/GSYH oranının düşüşü, ihracatın gerilemesi ve cari açığın baskısı, bize gösteriyor ki artık otoban sürüşün koşulları geçerli değil.

Ancak off-road'u asıl "intense" hale getirecek olan şok henüz tam olarak yaşanmadı: yapay zekânın üretim ve istihdam yapısını kökten değiştirmesi. Generatif AI (ChatGPT, Gemini, Claude gibi büyük dil modelleri) 2023'ten itibaren beyaz yakalı işleri — hukuk, muhasebe, yazılım, müşteri hizmetleri — doğrudan tehdit etmeye başladı. Agentic AI ise bir adım ötesine geçiyor: tek bir görev yapmak yerine, çoklu adımlı iş süreçlerini otonom yürüten yapay zekâ ajanları, orta düzey yönetim ve koordinasyon işlevlerini devralmaya aday. OECD'nin 2023 İstihdam Görünümü raporuna göre üye ülkelerde işlerin %27'si yüksek otomasyon riski altındadır; IMF ise yapay zekânın küresel işlerin yaklaşık %40'ını etkileyeceğini, gelişmiş ekonomilerde bu oranın %60'a çıkabileceğini öngörmektedir.

Bu dalgaya fiziksel üretim tarafından insansı robotlar eşlik ediyor. Tesla'nın Optimus'u, Boston Dynamics'in Atlas'ı, Figure AI ve Agility Robotics gibi firmalar, fabrika ve lojistik ortamlarında insanın yerini alabilecek genel amaçlı robotlar geliştiriyor. Eğer bu teknolojiler 2030'larda ölçeklenirse, off-road zemini bir kez daha değişecek: düşük maliyetli emek avantajına dayalı kalkınma modeli — ki Türkiye'nin imalat ihracatı hâlâ büyük ölçüde buna dayanıyor — geçerliliğini yitirecek. Ucuz işçilik artık rekabet avantajı değil, yapısal kırılganlık haline gelecek.

Mühendislik Ekonomisi perspektifinden bu üçlü dalga (generatif AI + agentic AI + insansı robotlar) off-road'un yeni engelleri değil, doğru okunan fırsat pencereleridir. Türkiye'nin genç, dijital okuryazar nüfusu ve mevcut yazılım/oyun ekosistemi (Dream Games, Trendyol, Peak Games mirası), AI çağında emek yoğun üretimden bilgi yoğun üretime geçişin altyapısını sağlayabilir — ancak bu geçiş kendiliğinden olmaz; eğitim müfredatının AI okuryazarlığıyla güncellenmesi, üniversitelerin AI araştırma kapasitesinin artırılması ve regülasyon mühendisliğinin bu alanı da kapsaması gerekir. Off-road çok daha yoğun olacak; mesele bu yoğunluğu kaos olarak mı yoksa ivme olarak mı deneyimleyeceğimiz.

## Yeni Tür(k) Kalkınma Modeli: Mühendislik Ekonomisi

Bu yeni off-road dönemde Türkiye'nin avantajı, mühendislik esnekliği olmalıdır. Buna dair başlıca özellikler olarak şunları sıralayabiliriz:

- Değişken zemine uygun üretim kapasitesi
- Genç ve teknik iş gücü
- Enerji, savunma ve dijital dönüşüm alanlarında hızlı adapte olabilen yapılar
- Hızlı karar verme yetisi
- Regülasyonları kullanarak teknoloji çekme becerisi

Bu nedenle Türkiye'nin izlemesi gereken rota, klasik finansal kalkınma değil, Mühendislik Ekonomisi temelli kalkınma olmalıdır. Yani ekonomi, hızı artırmak yerine zemine göre ayarlayarak ilerleyen bir "teknik strateji" ile yönetilmelidir. Kısacası, klasik ekonomi politikalarının ötesine geçerek ekonomik sistemi bir mühendislik problemi gibi ele almalıyız. Her değişken ölçülür, her risk modellenir, her kaynak optimize edilir. Büyüme artık yalnızca finansmanla değil, tasarım, inovasyon ve geri besleme döngüleriyle sağlanabilir. Enerjiden üretime, tarımdan dijital altyapıya kadar her alan, birer mühendislik sistemi olarak görülmelidir.

Bunun için yapısal ve taktiksel zemini kısa ve orta vadeli aksiyonlarla hazırlamalıyız.

### *1. Eski Paradan Yeni Para Çıkarmak*

İlk olarak ekonomik aktörlerimizi yenilemeliyiz. Burada önemli bir ince ayar yapmamız gerekiyor: Holding yapısı kendi başına bir sorun değildir. Nitekim Kore'nin çaebol'leri (Samsung, Hyundai, LG) de çok iş kollu, aile kontrolündeki dev yapılardı ve Kore mucizesinin omurgasını oluşturdular. Farkı yaratan yapı değil, yapının zorlandığı yöndü: Kore devleti çaebol'leri ihracata ve AR-GE'ye mecbur bıraktı, koruma önlemlerini ihracat performansına bağladı.

Samsung vakası, Kore sanayileşmesindeki daha geniş bir örüntüyü gözler önüne seriyor: Lee'nin (2019) "yan yol stratejileri" (detour strategies) dediği yaklaşımın bilinçli inşası. Koreli firmalar, olgun teknolojilerde yerleşik liderlerle doğrudan rekabet etmek yerine, mevcut oyuncuların avantajlarının en zayıf olduğu yeni teknolojik paradigmaları hedef aldılar. Yarı iletkenlerde DRAM'dan NAND flash belleğe geçiş böyle bir fırsat penceresi yarattı; ekran teknolojilerinde CRT'den LCD'ye ve ardından OLED'e geçiş bir başkasını açtı. Her seferinde Koreli firmalar yeni paradigmanın erken evresinde pazara girdiler, teknoloji henüz olgunlaşırken üretim kapasitesine devasa yatırımlar yaptılar ve Batılı ile Japon rakipler tepki veremeden maliyet liderliğini ele geçirdiler. Pazar girişini teknolojik paradigma değişiklikleriyle zamanlama stratejisi, bir sörfçünün dalgayı yakalamak için doğru anı kollamasına benzer — tam da Mühendislik Ekonomisi'nin savunduğu adaptif, fırsat-odaklı davranış biçimi.

Hyundai'nin otomotivdeki yörüngesi benzer bir kalıp izledi. 1970'lerde Ford araçlarının lisanslı montajcısı olarak başlayan Hyundai, kademeli olarak yerli tasarım ve mühendislik kapasitesini geliştirdi ve 1975'te İtalyan tasarımı ile Japon aktarma organları kullanan ilk tamamen Kore tasarımı otomobili Pony'yi piyasaya sürdü. 2000'li yıllara gelindiğinde Hyundai-Kia dünyanın beşinci büyük otomobil üreticisi haline geldi; 2020'lerde ise elektrikli araç ve hidrojen yakıt hücresi teknolojisinde öncü konumuna yükseldi. Lisanslı montajdan teknoloji liderliğine geçiş yaklaşık dört on yıl sürdü — Mühendislik Ekonomisi'nin öngördüğü nesiller arası kararlılıkla tutarlı bir zaman çizelgesi.

Türk holdingleri ise benzer yapısal avantaja sahip olmalarına rağmen bu zorlamayla karşılaşmadılar. Yabancı şirketler, holdingler üzerinden iç pazarı kontrol etti. Holdingler riski dağıtmak için farklı iş kollarına açıldı ve AR-GE ihtiyaçlarını yabancı ortaklarından karşıladıkları için, ülke içindeki üniversitelerin AR-GE kabiliyetleri yeterince kullanılmadı. Üniversiteler, yalnızca istihdam ve eğitim alanları dışında sınırlı şekilde değerlendirildi. Sorun dolayısıyla holdinglerin varlığı değil, AR-GE'den kopuk olmaları ve risk sermayesi ile bağlantısız kalmalarıdır.

Bu nedenle holdingler mevcut işleri yapmaya devam edebilir; ancak ana konuları dışındaki yatırımlarını sınırlandırmak ve bu kısmı teknoloji odaklı risk yatırım firmalarına yönlendirmelerini teşvik etmek gerekir. Kore modelinden alacağımız ders, çaebol'leri taklit etmek değil; sermayeyi yapısal olarak AR-GE ve ihracat performansına bağlamaktır. Bu yaklaşım, sermayeyi off-road'a uygun, esnek ve yenilikçi bir yapıya kanalize edecek ve ülke olarak kolektif şekilde her yıl ciddi miktarda (hedef olarak 20 milyar dolar mertebesinde) yatırımın risk yatırım firmalarında toplanmasını sağlayabilir. Bir anlamda eski sermaye, sistematik olarak yeni sermayeye dönüştürülecektir.

### *2. Elektroşok ile Hızlı Adaptasyon*

Türkiye'nin off-road ortamında hızla adaptasyon sağlaması için teknoloji devlerine ortaklık stratejisi izlemeliyiz. Risk yatırım firmalarında biriken sermayenin anlamlı bir kısmını seçilmiş teknoloji devlerine yatırarak onların hissedarı olmalıyız. Bu şekilde her yıl "teknoloji şokları" yaratarak teknoloji trenine kanca atabilir ve iç kamuoyunun farkındalığını artırabiliriz.

Hangi teknoloji alanının veya şirketin seçileceği, devletin ana odak noktaları ile Türkiye Varlık Fonu'nun organizasyonunda risk yatırım firmalarının kollektif çalışması sonucu belirlenir. Örneğin, devlet insansı robotlar temasını seçebilir; diğer paydaşlar ise hangi şirkete yatırım yapılacağını kararlaştırır. Bu alanlar gelişmiş ülkelerin uzun vadeli yatırım yaptıkları olgun teknolojilerden ziyade, kısa vadeli fikri haklar yeni şekillenen yan yollardan seçilmeli ve gelişmekte olan bir ülke olarak aradan çıkmanın yollarını aramalıyız. Bunun için teknoloji fırsat penceresi önemlidir.

Bu strateji, şirketin AR-GE ihtiyaçlarını kendi üniversitelerimizden karşılamasını sağlayacak bir sinerjiyi başlatır. Her yatırımın arkasına bir grup üniversitemizi de bağlayabiliriz. Aynı zamanda şirketin istihdam yapısında üniversitelerimizin mezunları yer alır ve yan tamamlayıcı ürünleri ülkemizde geliştirilebilir. Mevcut sanayimiz de bu yönlendirmeler doğrultusunda şekillenir. Bu süreci her yıl tekrarlayarak sürekli bir teknoloji ve inovasyon döngüsü yaratabiliriz.

Örneğin, 2019 sonunda Tesla hissedarı olsaydık, bölünme sonrası düzeltilmiş fiyatla 10–15 kat civarı bir değer artışı yaşayabilirdik; NVIDIA'da ise 2020 başından bu yana benzer büyüklükte bir çarpan söz konusudur. Ancak bu örneklerin retrospektif olarak seçildiğini unutmamak gerekir: aynı dönemde Peloton, Zoom veya WeWork gibi şirketlere yatırılan sermaye ciddi kayba uğramıştır. Doğru model tek şirket seçmek değil, 10–15 stratejik alana yayılmış bir portföyün beklenen getirisidir. Böylece hem ülke olarak fayda sağlarız hem de mevcut sermaye sahipleri kazanır; ama tek bir "Tesla'yı yakalama" stratejisi ciddi bir geriye dönük yanılsama tuzağıdır.

### *3. Regülasyon Mühendisliği ile Teknoloji Parası Çekmek*

Off-road ekonomide başarı, yalnızca sermayenin değil, regülasyonun mühendisliğiyle mümkündür. Çünkü teknolojinin kendisi kadar hızlı ve esnek olmayan hiçbir hukuk, yönetmelik veya izin sistemi, yatırımı değil, tereddüdü büyütür. Mühendislik Ekonomisi yaklaşımı bu noktada "kuralları gevşetmek" değil, "kuralları akıllandırmak" üzerine kurulmalıdır. Bugün ülke olarak hâlâ birçok alanda AB'nin takibinde kalarak regülasyonları daha katı bir şekilde takip ediyoruz. Bunun için diğer ülkelerde olmayan karar verme hızımızı öne çıkarmalıyız.

#### *3.1 Regülasyonun Hızı: Fren mi, Süspansiyon mu?*

Klasik ekonomi dönemlerinde regülasyon, fren işlevi görürdü — aşırılığı engellerdi. Off-road döneminde ise regülasyonun görevi değişmiştir: artık süspansiyon görevi görmelidir. Yani darbeleri yumuşatmalı, sistemin esnekliğini korumalıdır. Teknoloji hızlandıkça, regülasyonun da "hız eşlemesi" gerekir. Aksi halde, inovasyonun tekeri regülasyon taşına çarpar ve araç devrilir.

#### *3.2 Alan Açan Regülasyon Modeli (Sandbox Yaklaşımı)*

Dünya genelinde özellikle yapay zekâ, biyoteknoloji ve fintech alanlarında "regülasyon sandbox" adı verilen esnek deneme alanları kullanılmaktadır. Bu sistemde girişimler sınırlı ölçekte ama gerçek ortamda ürünlerini test eder, devlet ise regülasyon riskini gözlemler. Mühendislik Ekonomisi yaklaşımında:

- İlk aşamada teknoloji firmalarına "geçici test izni" verilir.
- İkinci aşamada bu testlerin çıktısına göre kalıcı düzenlemeler yapılır.
- Son aşamada başarılı örnekler genelleştirilir.

Böylece regülasyon, inovasyonu kısıtlamaz, inovasyonun rehberine dönüşür.

#### *3.3 Veri ve Sermaye Arasında Köprü Kurmak*

Yeni çağda veri, sermayeden bile daha değerlidir. Bu nedenle regülasyonların iki hedefi aynı anda gözetmesi gerekir:

1. Veriyi korumak (vatandaş güvenliği için),

2. Veriyi akışkan hale getirmek (yatırım ve inovasyon için).

Bu denge kurulmadan ne yabancı teknoloji fonları gelir ne de yerli girişimler küresel pazara açılabilir. Mühendislik Ekonomisi yaklaşımı bu dengeyi "kontrollü geçirgenlik" olarak tanımlar: sistem ne tamamen kapalı ne de tamamen açık olur.

### *3.4 Regülasyon Mühendisliğinde Ölçüt: Hız + Güven + Uyarlanabilirlik*

Bir ülkenin teknolojiye uygunluğu artık bu üç ölçütle belirleniyor:

- Hız: İzin süreçleri, yatırım döngülerine uyumlu mu?
- Güven: Yatırımcıya ve vatandaşa aynı anda öngörülebilirlik sağlıyor mu?
- Uyarlanabilirlik: Yeni teknolojiler çıktığında sistem kendini yenileyebiliyor mu?

Türkiye bu üç unsuru aynı anda geliştirdiği anda hem finansal fonlar hem teknoloji fonları için "denge noktası" haline gelir.

### *3.5 Devletin Yeni Rolü: Düzenleyici değil, Eş Tasarımcı*

Mühendislik Ekonomisi yaklaşımında devletin rolünü yeniden tanımlayabiliriz. Artık devlet yalnızca "düzenleyen" değil, "ortak tasarlayan" aktör olabilir. Yani teknoloji firmalarıyla birlikte *co-regulation* denilen sistemde çalışır. Bu sistemde:

- Şirketler, ürün geliştirmeden önce regülasyon otoritesiyle masaya oturur.
- Regülasyon, üretimin önünde değil, üretimin içinde yer alır.
- Denetim sonradan değil, eşzamanlı olarak yapılır.

### *3.6 Regülasyonun Diplomasisi: Global Uyumun Gücü*

Teknoloji artık sınır tanımadığı için, ulusal regülasyonlar da izole kalamaz. AB'nin yapay zekâ yasası (AI-Act) ya da ABD'nin veri paylaşım protokolleri, yatırım kararlarını doğrudan etkiliyor. Türkiye'nin Mühendislik Ekonomisi yaklaşımı, küresel regülasyon ekosistemleriyle senkronizasyonu hedeflemelidir. Bu sayede Türkiye, "uyum sağlayan ülke" değil, uyumu yönlendiren ülke haline gelir.

Regülasyon mühendisliği, sadece bir hukuki reform değil, yatırım ortamının fiziksel konfor ayarıdır — süspansiyonu doğru ayarlanmış bir ekonomi, en bozuk yolda bile hızını korur.

***4. ABD ve Çin Teknoloji Geriliminden Beslenmek***

Küresel teknoloji rekabeti, Türkiye için bir fırsat alanı yaratabilir. ABD ve Çin arasındaki gerilim, bazı kritik teknolojilerin ve üretim hatlarının yeniden dağıtılmasını zorunlu kılıyor. Türkiye, bu durumu stratejik olarak kullanabilir. Örneğin, ABD'de Intel veya Apple gibi teknoloji devleri, tedarik zincirlerini çeşitlendirmek veya Çin'e bağımlılıklarını azaltmak istiyor. Bu durumda Türkiye, üretim, AR-GE veya pilot test alanları sunarak bu şirketlerle ortaklık kurabilir.

Benzer şekilde Çin'in genişleyen teknoloji ve üretim kapasitesi, Türkiye'ye ortak üretim veya tedarik zinciri entegrasyonu fırsatı sağlayabilir. Örneğin, Çin'de üretilen elektronik bileşenlerin bir kısmı Türkiye'de monte edilebilir, yerli yazılım ve mühendislik iş gücü ile birleştirilerek hem Türkiye ekonomisine değer katılır hem de küresel tedarik zincirinde Türkiye stratejik bir rol üstlenir.

Bu yaklaşım, Türkiye'nin off-road ortamında hızla adaptasyon sağlamasını ve teknoloji elektroşoklarını ülke içine çekmesini mümkün kılar. Her yeni teknoloji devrimi, risk yatırım firmaları ve devletin ortak aksiyonu ile belirlenen alanlarda test edilerek ülkeye kalıcı bir değer yaratır. Bundan dolayı Türkiye'nin ABD ve Çin'deki büyükelçileri teknoloji odaklı profillere sahip olmalı; Teknoloji Büyükelçiliği daha aktif çalışmalıdır. Fırsat pencerelerinin hızla açılıp kapandığı bu dönemde hamle yapmak belirleyicidir. Örneğin:

***1. AB Yapay Zekâ Yasası (AI Act) Mültecileri İçin "Sandbox"***

- Küresel Durum (Fren): Avrupa Birliği, Yapay Zekâ Yasası (AI Act) ile inovasyonu sıkı kurallara bağladı. Birçok Avrupalı AI girişimi (startup), ağır bürokrasi ve uyum maliyetleri nedeniyle ürün geliştirmekte zorlanıyor.
- Türkiye Fırsatı (Gaz): "Regülasyon Sandbox" (Güvenli Deneme Alanı) devreye alınabilir. Türkiye, "Gelin algoritmalarınızı burada test edin, veriyi burada işleyin, dünyaya buradan açılın" diyebilir.
- Vaat: AB standartlarına uyumlu ancak "geliştirme aşamasında esnek" bir yasal liman sağlamak.

***2. Almanya'nın Enerji Krizi ve "Sanayi Göçü"***

- Küresel Durum (Fren): Almanya ve Orta Avrupa'da enerji maliyetleri çok yüksek olduğu için ağır sanayi ve veri merkezleri (Data Centers) rekabetçiliğini kaybediyor.
- Türkiye Fırsatı (Gaz): Türkiye'nin yenilenebilir enerji potansiyeli kullanılarak, enerji yoğun teknoloji şirketlerine (özellikle veri merkezleri) "Yeşil Enerji Teknoloji Bölgeleri" sunulabilir.
- Vaat: "Fabrikanızı buraya taşıyın, size 10 yıl boyunca sabit ve düşük maliyetli yenilenebilir enerji garantisi verelim".

***3. Oyun Sektöründeki "Çin Kısıtlamaları"***

- Küresel Durum (Fren): Çin, oyun sektörü üzerinde ciddi kısıtlamalar (oyun süresi limitleri, lisans durdurmaları) uyguluyor. Bu durum, dünyanın en yetenekli oyun geliştiricilerini ve stüdyolarını baskı altına alıyor.
- Türkiye Fırsatı (Gaz): Türkiye halihazırda bir oyun (gaming) hub'ı. Bu Çinli ekiplere ve bağımsız stüdyolara özel teşvikler verilebilir.
- Vaat: Teknoloji Büyükelçiliği mantığıyla, Şanghay veya Pekin'deki yetenekli oyun tasarımcılarına "Yaratıcı vize" ve İstanbul'da ortak çalışma alanları sağlanabilir.

***4. Kripto ve Web3 için "Netlik Limanı" (ABD SEC Baskısı)***

- Küresel Durum (Fren): ABD Menkul Kıymetler ve Borsa Komisyonu (SEC), kripto firmalarına karşı belirsiz ve sert bir tutum sergiliyor. Bu durum Web3 geliştiricilerini kaçırıyor.
- Türkiye Fırsatı (Gaz): "Kuralları gevşetmek değil, akıllandırmak" ilkesi burada uygulanabilir.
- Vaat: "Yasakçı değil, tanımlayıcı" bir dijital varlık yasası ile Web3 geliştiricilerine hukuki öngörülebilirlik sunmak.

*5. Biyoteknoloji ve "Hızlı Klinik Deney" (FDA Yavaşlığı)*

- Küresel Durum (Fren): ABD'de FDA onay süreçleri veya Avrupa'daki klinik deney süreçleri yıllar sürüyor ve çok maliyetli. Bu durum "Longevity" (uzun yaşam) ve biyoteknoloji araştırmalarını yavaşlatıyor.
- Türkiye Fırsatı (Gaz): Türkiye'nin güçlü sağlık altyapısı ve hızlı karar verme yetisi kullanılarak, etik kurallardan taviz vermeden bürokrasisi hızlandırılmış "Biyoteknoloji Serbest Bölgeleri" kurulabilir.
- Vaat: Araştırmacılara, "Molekülünüzü buldunuz, faz çalışmalarını 3 yılda değil, 1 yılda tamamlayabileceğiniz hızlı ve güvenli altyapı bizde" denilebilir.

*6. ABD Araştırma Fonu Kesintileri*

- Küresel Durum (Fren): Son dönemde ABD üniversitelerine sağlanan federal ödeneklerin kesilmesi veya belirli araştırma alanlarının fonsuz bırakılması, nitelikli araştırmacıları yeni kaynak ve laboratuvar arayışına itmektedir.
- Türkiye Fırsatı (Gaz): Bu araştırmacılar, henüz yerleşmemişken seçilmiş Türk üniversitelerine davet edilebilir. Bu hamle, sadece bilgi transferi sağlamakla kalmaz, Türkiye'nin akademik ve entelektüel imajını küresel ölçekte yukarı çeker.
- Vaat: "Araştırmanız fon kesintisine kurban gitmesin; gelin laboratuvarınızı ve ödeneğinizi Türkiye'de biz sağlayalım, çalışmalarınıza kesintisiz devam edin."

*7. Apple ve Tedarik Zinciri Göçü (China Plus One Stratejisi)*

- Küresel Durum (Fren): Jeopolitik gerilimler ve tedarik zinciri riskleri nedeniyle Apple (ve Foxconn gibi tedarikçileri), üretimlerinin büyük kısmını Çin dışına çıkarmaya çalışıyor. Ancak Hindistan ve Vietnam'daki altyapı/kalite sorunları bu geçişi zorlaştırıyor ve yavaşlatıyor.

- Türkiye Fırsatı (Gaz): Türkiye, Gümrük Birliği avantajı, gelişmiş lojistik altyapısı ve kalifiye montaj sanayisi ile bu arayışta "en güvenli ve hızlı liman" olarak öne çıkabilir. Sadece montaj değil, parça üretimi için de devreye girilebilir.
- Vaat: "Çin'deki riskinizi dağıtmak için macera aramayın; gelin üretiminizin stratejik bir kısmını (özellikle Avrupa pazarı için olanı) Türkiye'deki 'Teknoloji Üretim Bölgeleri'ne kaydırın."

| ***Karşılaştırma Notu: Dubai Krizi ve İstanbul Finans Merkezi — Fırsat Penceresi Örneği*** | | |
|---|---|---|
| | ***Dubai (Mevcut Durum)*** | ***İstanbul Finans Merkezi (TR Hamlesi)*** |
| ***Tetikleyici*** | ***İran savaşı Dubai'nin bölgesel finans merkezi konumunu zayıflattı; yatırımcılar alternatif arıyor*** | ***Türkiye, boşluğu fırsat penceresine çevirmek için 15 maddelik yasa teklifi ile İFM teşviklerini genişletti (Nisan 2026)*** |
| ***Teşvik Yapısı*** | ***Sıfır vergi ama dar kapsamlı, finansa odaklı*** | ***Kurumlar vergisi indirimi %50→%100, finansal hizmet ihracatı teşvikleri 2047'ye kadar, harç muafiyeti 20 yıl, nitelikli personel gelir vergisi istisnası*** |
| ***Coğrafi Avantaj*** | ***4 saatlik uçuşla Körfez ve Güney Asya*** | ***4 saatlik uçuşla 1,3 milyar kişi — hem Avrupa hem Körfez hem Orta Asya*** |
| ***ME Perspektifi*** | ***Otoban dönemi finans merkezi: statik vergi avantajı, pasif bekleme*** | ***Off-road hamlesi: krizi fırsata çeviren, hızlı regülasyon mühendisliği ile proaktif pozisyon alma*** |

### *5. Finansal Fonlardan Ziyade Teknoloji Fonlarını Hedeflemek*

Eğer ekonomik düzenimiz bir otoban olsaydı, kısa vadeli finansal fonların ilgisini çekebilmek için iç mekanizmalarımızın düzgünlüğü ve güvenilirliği yeterli olurdu. Ancak biz bugün off-road bir yoldayız; bu nedenle teknoloji fonlarını çekmek için ortaya koyacağımız vizyon belirleyici olacaktır.

Teknolojinin gelişip yer bulabilmesi için uygun bir ekosistemin önemi her zamankinden daha fazla. Günümüzde teknoloji fonları, finansal fonların önüne geçmiş durumda. Bu yeni düzende şirketlerle stratejik iş birlikleri öne çıkıyor. Nitekim NVIDIA, 100 milyar doların üzerinde yatırım yapma kararı alarak bu dönüşümün ölçeğini bir kez daha gösterdi. Teknoloji fonlarında başarı sağladığımızda, konvansiyonel finansal fonlar da doğal olarak kendiliğinden gelecektir.

#### *5.1 Otoban Zamanı Yabancı Finansal Fonlar (2000–2008 Türkiye'si gibi)*

- Ne isterler: Hızlı ve güvenli getiri; düşük risk.
- Ne verirler: Sermaye, bazen teknoloji transferi ve yönetim tecrübesi.
- Örnekler: Telekom sektörü (Vodafone Türkiye), enerji sektörü (BP, Shell), sanayi (Ford, Toyota) — sermaye ve üretim katkısı sağladılar; ancak stratejik teknoloji transferi ve AR-GE açısından etkileri sınırlı kaldı.

#### *5.2 Off-road Zamanı Yabancı Yatırımcıları (2020 sonrası)*

- Ne isterler: Yüksek riskli ama stratejik fırsatlar; teknoloji ve pazar avantajı.
- Ne verirler: Sermaye + ileri teknoloji + AR-GE + know-how + küresel bağlantılar; pasif yatırım değil, aktif iş birliği beklerler.
- Örnek (Tesla): Türkiye stratejik hisse alsaydı, sermaye yanında batarya teknolojisi, elektrikli araç AR-GE'si, fabrika planlama ve üretim teknolojisi transferini de elde edebilirdi.
- Örnek (NVIDIA / Intel / AMD): GPU ve yapay zekâ teknolojileri yerel AR-GE ve üretim altyapısıyla entegre edilebilir, üniversitelerle güçlü iş birlikleri oluşturabilir.
- Örnek (SpaceX / Blue Origin): Uzay teknolojileri ve yan endüstriler Türkiye'ye çekilebilir.

- Örnek (UBER): Pazara giriş karşılığı istihdam yerine Türkiye Varlık Fonu'na hisse talep edilebilir. Bu modelin hukuki çerçevesi, DTÖ ve Gümrük Birliği taahhütleri ile yatırımcı-devlet tahkim sistemi dikkate alınarak özenle tasarlanmalıdır; aksi halde uluslararası hukuki itirazlara açık kalır.

Hazine ve Maliye Bakanlığı, Maliye ve Ekonomi olarak ikiye bölünebilir ve Ekonomi Bakanlığı içinde "Teknoloji Sermaye Dairesi" kurulmalı, yatırım kararları mühendislik kapasitesiyle uyumlu hâle getirilmeli ve ekonomik büyüme "üretim + teknoloji + risk sermayesi" eksenine taşınmalıdır. Bu yaklaşım, sermaye ve teknoloji üretimini birleştiren, off-road koşullarına uygun, stratejik esnekliği yüksek bir ekonomik sistem yaratacaktır.

***6. Yüksek Enflasyon ve Para Politikası Uyumsuzluğu***

Klasik ekonomi döneminde para politikası, otoban üzerindeki bir aracın çekiş kontrol sistemine benzer: zemin düz olduğu için faiz yükseltip indirerek hızı ayarlamak yeterlidir. Ancak off-road dönemde zemin düz değildir; ani kaymalar, dış şoklar ve yapısal dengesizlikler bu mekanizmayı tek başına yetersiz kılar. Faiz, bu yeni dönemde tek başına "gaz pedalı" değildir; sistemin çekişini koruyan birçok kontrol hattından sadece biridir.

Mühendislik Ekonomisi yaklaşımı, bu noktada ekonomiyi dinamik bir sistem olarak ele alır. Enflasyon, sadece fiyatların artışı değil, sistemin farklı bileşenleri arasındaki çekiş kaybının bir göstergesidir. Bu nedenle, zemine göre uyum sağlayabilen, çok değişkenli ve geri beslemeli bir para politikası önerir.

Bu modelde Hazine ve Merkez Bankası kararları, statik takvimlere göre değil, model tabanlı kontrol sistemleri üzerinden yürütülür. Ekonomi, sürekli ölçüm yapan sensörlerle (fiyat endeksleri, enerji maliyetleri, kredi genişlemesi, döviz giriş-çıkışları gibi) izlenir. Kararlar, bu sensörlerden gelen verilere göre otomatik ve kademeli olarak ayarlanır. Örneğin, Singapur ve Güney Kore'nin uyguladığı dinamik faiz ve likidite yönetim sistemleri, bu anlayışın başarılı örnekleridir. Türkiye'de de benzer şekilde faiz oranları, zorunlu karşılıklar veya rezerv yönetimi, insan takvimiyle değil, veri tabanlı algoritmik sistemlerle güncellenebilir. Böylece faiz politikası, politik bir araç değil, teknik bir parametreye dönüşür.

Ayrıca, döviz kuru ve likidite birer "kontrol değişkeni" olarak ele alınır. Döviz kuru, artık müdahale edilen bir çıktı değil, zeminin eğimini ölçen bir sensördür. Piyasa oynaklığı belirli eşikleri aştığında, sistem otomatik olarak swap, rezerv veya tahvil alım-satım araçlarını devreye sokabilir. Böylece para politikası manuel müdahaleye değil, yarı otonom mühendislik sistemine dayanır.

Son olarak, enflasyonun kaynağına inen çoklu sensör sistemleri kurulmalıdır. Tarımsal arz eksikliği, enerji maliyet artışı veya kredi genişlemesi gibi farklı nedenler, ayrı kontrol devreleriyle yönetilir. Tarımsal enflasyon stok optimizasyonuyla, enerji enflasyonu ithalat vadeleriyle, talep enflasyonu kredi mekanizmalarıyla dengelenir.

Bu yaklaşım, klasik ekonominin "tek araç – tek hedef" anlayışını geride bırakır. Artık ekonomi bir mühendislik sistemi gibi işler:

- Faiz, çekiş kontrolüdür — tek başına gaz veya fren değil, zemine göre itkiyi koruyan bir ayardır.
- Döviz kuru, hedef değil, pusuladır.
- Enflasyon, arızadan çok sinyaldir.

Sonuç olarak, Mühendislik Ekonomisi yaklaşımı, Türkiye'nin para politikasını statik olmaktan çıkarıp, zemine göre sürekli ayarlanan bir "akıllı süspansiyon sistemi"ne dönüştürür. Bu sistem, off-road şartlarında sarsıntıyı azaltır, yön kaybını önler ve ekonomiyi kontrollü bir biçimde hedefe taşır.

**Karşılaştırma Notu: Mevcut Para Politikası Çerçevesi vs. Mühendislik Ekonomisi**

Mevcut para politikası çerçevesi bu ihtiyacın ne kadar gerisinde kaldığını somut biçimde göstermektedir. Politika faizi önce %50'den %42,5'e indirilmiş, ardından Nisan 2025'te piyasa volatilitesi nedeniyle 350 baz puanlık artışla %46'ya geri çekilmiş, sonra tekrar beş ardışık indirimle %37'ye düşürülmüş ve bölgesel savaşın tetiklediği enerji şokuyla iki toplantı boyunca sabit tutulmuştur — klasik "sıkılaştır-gevşet" zigzagının off-road zeminde tekrarı. Nisan 2026 itibarıyla yıllık enflasyon %32 seviyesindeyken orta vadeli hedef %5'tir; aradaki uçurum, tek araç–tek hedef yaklaşımının sayısal fotoğrafıdır. Mevcut sistem jeopolitik şokların yapısal niteliğini, enflasyonun çok kaynaklı doğasını ve enerji fiyatlarının cari dengeye baskısını doğru teşhis etmektedir; ancak önerdiği reçete hep aynıdır: faizi yükselterek soğut, indirerek ısıt. Mühendislik Ekonomisi ise enflasyonu düşürülmesi gereken bir sorun değil, çekiş kaybının sinyali olarak okur; bütçe açığını zafiyet değil, yönü doğru olduğunda stratejik bir yatırım aracı olarak görür; ve sabit takvimli kararlar yerine veriye dayalı algoritmik ayarlama önerir. Kısacası, mevcut çerçeve motoru yalnızca gaz-fren pedalıyla (faiz) yönetmeye çalışmakta; oysa off-road zeminde süspansiyon (sektörel regülasyon), vites (üretim yapısı dönüşümü) ve çekiş kontrol sistemi (hedefe yönelik kredi yönlendirmesi) devrede olmadığında aracın tam performansına ulaşması güçleşir.

### *7. Bütçe Açıkları ve Borçlanma*

Klasik ekonomi yaklaşımında bütçe, bir muhasebe defteridir: gelirler bir tarafa, giderler diğer tarafa yazılır ve denge aranır. Ancak off-road dönemlerinde bu yöntem yetersiz kalır; çünkü zemin sürekli değişir, gelir ve gider eğrileri öngörülemez biçimde dalgalanır.

Mühendislik Ekonomisi, bütçeyi statik bir defter değil, enerji akışı yöneten bir sistem olarak görür. Burada asıl ayrım, açığın varlığı değil, açığın yönüdür. Türkiye'nin bugün yaşadığı pro-siklik açık (yani büyümede de büyüyen açık) bir arızadır; ancak karşı-konjonktürel, üretken yatırıma giden bir açık sistemin adaptasyon kapasitesidir. Dolayısıyla Mühendislik Ekonomisi "açık iyidir" demez; açığın hangi tür olduğunu ayırt eder.

Bu yaklaşımda, devletin borçlanma politikası da klasik finansal muhafazakârlığın ötesine geçer. Borç, yalnızca cari harcamaları finanse eden bir yük değil, dönüşüm yatırımlarına güç sağlayan bir tork olarak tasarlanır. Kritik olan, borcun yönü ve geri dönüş katsayısıdır.

- Eğer borç, üretken kapasiteyi (örneğin yarı iletken, batarya, yeşil enerji veya yapay zekâ yatırımlarını) artıran alanlara yöneliyorsa, bu "pozitif geri beslemeli borç"tur.

- Eğer borç, cari harcama veya kısa vadeli ithalat finansmanına gidiyorsa, bu "enerji kaçağı"dır.

Bu ayrım, Mühendislik Ekonomisi'nde statik mali disiplin yerine dinamik mali verim kavramını öne çıkarmasının nedenidir. Disiplin ile verimlilik birbirine zıt değildir; verimli bir bütçe zaten disiplinlidir. Fakat disiplin tek başına, üretken olmayan bir bütçeyi hayırlı yapmaz.

Mühendislik Ekonomisi, borçlanma portföyünü mühendislik dengesiyle optimize etmeyi önerir. İç borç – dış borç dengesi, yalnızca faiz farkına göre değil, döviz likiditesi, jeopolitik riskler ve üretim zinciri bağımlılıklarına göre hesaplanır. Böylece devlet, borçlarını statik bir faiz maliyetiyle değil, dinamik risk modeliyle yönetir. Örneğin, Japonya'nın iç tasarruflara dayalı borçlanma modeli veya Güney Kore'nin proje bazlı uluslararası tahvil ihracı, Mühendislik Ekonomisi anlayışına yakındır. Türkiye de benzer biçimde, klasik borçlanma yerine yatırım temalı devlet tahvilleri (örneğin "AI Bond", "Green Transition Bond", "Defense Innovation Bond") çıkarabilir.

Bu modelde, borçlanmanın zamanlaması da statik değildir. Klasik ekonomi yıllık bütçe döngüsüne bağlı kalırken, Mühendislik Ekonomisi yaklaşımı "modüler bütçe" fikrini benimser. Her çeyrekte, harcama akışı performansa göre yeniden ayarlanır; düşük performanslı kalemlerin finansmanı kesilir, yüksek çarpanlı projeler hızlandırılır.

Sonuçta, Mühendislik Ekonomisi'nin borçlanma modeli fiziksel bir denge kurar:

- Harcama bir güçtür, borç bir torktur, bütçe ise moment koludur.
- Denge, sıfır açığa değil, üretken yatırımın maksimum ivmesine göre ayarlanır.
- Hedef, istikrar ile verimli hızlanmanın birlikte sağlanmasıdır.

Bu bakış açısıyla Türkiye, klasik "borç-faiz-enflasyon" üçgeninin içinde sıkışmak yerine, mühendislik temelli "yatırım-verimlilik-teknoloji" döngüsünü inşa edebilir.

### *8. Döviz Kuru ve Dış Denge: Denge Motorunun Mühendisliği*

Klasik ekonomi döviz kurunu bir sonuç olarak görür: arz-talep dengesiyle belirlenen, dış ticaret açığı veya sermaye girişiyle hareket eden bir gösterge. Oysa Mühendislik Ekonomisi dövizi bir denge motorunun dönüş hızı olarak ele alır. Bu motor, ekonominin iç enerjisini dış dünyayla uyumlu bir dönme hızında tutar.

Bu bakışla döviz kuru, yalnızca ticaretin değil, ülkenin teknolojik yoğunluğu, üretim derinliği ve veri egemenliğinin bir yansımasıdır. İthalatın enerji yoğun, ihracatın ise düşük katma değerli olduğu bir ekonomi, torku dışa kaçan bir motora benzer: her hızlanmada enerji kaybeder.

Mühendislik Ekonomisi yaklaşımında döviz dengesi üç eksende yeniden tanımlanır:

1. Akış Mühendisliği (Flow Engineering): Döviz, sadece para değil, enerji akışıdır. Amaç, döviz girişini artırmaktan çok akış kayıplarını azaltmaktır. Bu da ithalat bağımlılığını azaltmakla değil, ithalatın içselleştirilmesiyle olur. Örneğin, batarya üretiminde kullanılan lityumun madencilikten kimyaya, yazılıma kadar yerli ekosistem içinde kalması, döviz çıkışını azaltmakla kalmaz; sistemin "akış viskozitesini" düşürür.

2. Denge Kontrolü (Balance Control): Döviz kuru, serbest bırakılmış bir değişken değil, geri besleme ile kontrol edilen bir parametre olmalıdır. Mühendislik Ekonomisi burada PID kontrol mantığını önerir: P (Proportional) — kısa vadeli müdahale (MB rezervi, faiz tepkisi); I (Integral) — uzun dönem ortalama denge (yapısal ihracat kapasitesi); D (Derivative) — şoklara tepki hızı (finansal esneklik). Bu üç kontrol hattı senkronize çalıştığında, döviz kuru doğal salınımlar yapar ama sistem rezonansa girmez.

3. Enerji Yoğunluğu ve Kur Hassasiyeti: Her sektörün dövize duyarlılığı farklıdır (örneğin otomotivde yüksek, yazılımda düşük ithal girdi oranı). Mühendislik Ekonomisi, makro kur politikasını tüm ekonomiye eşit uygulamak yerine sektörel kur modülasyonu önerir. Yazılım ihracatına düşük efektif kur, enerji ithalatına yüksek efektif kur uygulanabilir. Böylece sistem, enerji bağımlılığını azaltırken bilişsel ihracatı teşvik eder.

Bu yaklaşımda "dış denge" artık yalnızca ihracat-ithalat farkı değildir; ülkenin bilgi, veri ve inovasyon akışıyla ölçülen net enerji dengesidir. Bir ülke yazılım, tasarım veya yapay zekâ hizmeti ihraç ediyorsa, bu fiziksel mallardan daha yüksek "döviz enerjisi" taşır. Dolayısıyla Mühendislik Ekonomisi, dış dengeyi mal akışı değil, entropi dengesi olarak ölçer: düşük entropili ihracat (teknoloji, marka, patent) > yüksek entropili ithalat (hammadde, enerji).

Bu bakış, Türkiye'nin döviz politikasında yeni bir paradigma sunar. Döviz kuru artık piyasanın tahmin ettiği bir sonuç değil, ülkenin teknolojik momentumunun kontrol koludur. Tıpkı bir mühendislik motorunda olduğu gibi, amaç dönüş hızını sabit tutmak değil, ivmeyi kontrol altında artırmaktır.

### ***9. Faiz Oranı ve Sermaye Akışları***

Klasik ekonomi, faiz oranını sermayenin fiyatı olarak görür: paranın zaman değeri, risk primi ve enflasyon beklentisinin toplamı. Oysa Mühendislik Ekonomisi açısından faiz, yalnızca fiyat değil, sistemdeki potansiyel farkını gösteren bir enerji gradyanıdır. Bu bakış, Bölüm 6'daki çekiş kontrolü metaforunu tamamlar: faiz, gaz pedalı gibi tek başına hızı belirleyen bir kol değil, motorun sıkıştırma oranı, yakıt karışımı ve ateşleme sistemiyle birlikte çalışan bir çekiş-kontrol bileşenidir.

#### ***9.1 Potansiyel Farkı ≠ Sermaye Girişi***

Faiz oranı yükseldiğinde klasik beklenti, dışarıdan sermaye akışı olur yönündedir. Mühendislik Ekonomisi yaklaşımında ise bu ancak sistemin iç direnci düşükse gerçekleşir. Yani bürokrasi, belirsizlik, hukuki risk, siyasi gürültü gibi "akım direnci" yüksekse; yüksek faiz bile sermaye çekmez. Bu durumda sistem, yüksek voltaj altında ısınır ama çalışmaz — enerji (sermaye) harcanır, üretim artmaz. Bu yüzden yüksek faiz, yönlendirilmediği takdirde bir "soğutma mekanizması" değil, termal kaçak riskidir.

#### ***9.2 Sermaye Akışının Akışkan Dinamiği***

Faiz farkı, iki ekonomi arasındaki basınç farkı gibi düşünülebilir. Türkiye'nin teknolojik ve yapısal potansiyeli yüksek olsa da, güven endeksi düşükse, toplam potansiyel etkisi dengelenir ve sermaye çekimi sınırlı kalır. Gerçek sermaye akışını sağlamak için, faiz farkının sürdürülebilir bir vizyon farkıyla desteklenmesi gerekir.

### *9.3 Yapısal Kapasitör: Teknoloji ve Güven Birikimi*

Bir mühendislik sisteminde kapasitör, enerjiyi depolar ve şokları yumuşatır. Ekonomideki karşılığı, teknolojik kapasite ve toplumsal güven birikimidir. Eğer ekonomi kısa vadeli sermayeye bağımlıysa, bu kapasitör küçüktür — dışarıdan gelen şoklar sistem voltajını hızla değiştirir. Mühendislik Ekonomisi yaklaşımı, faiz politikasını yalnızca Merkez Bankası'nın değil, inovasyon ve güven politikalarının bir fonksiyonu olarak ele alır.

### *9.4 Sermaye Akışının Rezonans Noktası*

Her sistemin bir doğal frekansı vardır; ekonomi de bundan farklı değildir. Faiz oranı çok sık veya ani değiştiğinde, sistem kendi doğal frekansında titreşmeye başlar. Bu, yatırım kararlarını dondurur, finansal varlıkları aşırı oynak hale getirir. Mühendislik Ekonomisi bakışı, "faiz politikası"nı bir titreşim sönümleme problemi olarak ele alır.

### *9.5 Finansal Fonlar – Teknoloji Fonları Ayrımı*

Off-road ekonomilerde faiz farkı, artık tek başına yatırım çekme aracı değildir. Finansal fonlar potansiyel farkı izler; teknoloji fonları ise momentum farkına bakar. Yani kısa vadeli getiri değil, uzun vadeli yön önemlidir. Eğer sistemin dönüş yönü teknolojiye dönükse, düşük faiz bile sermaye çeker — çünkü faiz farkı yerine vizyon farkı devreye girer.

Özetle, faiz oranı para politikasında direksiyon veya gaz pedalı değil, sistemin çekişini koruyan bir kontrol bileşenidir. Gazın işe yaraması için motorun sıkıştırma oranı, yakıt karışımı ve ateşleme sistemi uyumlu olmalıdır. Faiz tek başına ekonomiyi çalıştırmaz; sistemin mühendisliğine bağlı olarak ya itki üretir ya da yakıt israfına yol açar.

## *10. Sermaye Birikimi ve Yatırım Motorunun Tasarımı*

Ekonomiyi sadece bir finansal sistem olarak değil, enerji dönüşüm motoru olarak görmek gerekir. Mühendislik Ekonomisi yaklaşımı, sermayeyi "yakıt", yatırımı "mekanik dönüşüm", teknolojiyi ise verim artırıcı türbin olarak ele alır. Bu yaklaşımda amaç, yalnızca sermaye biriktirmek değil, onu doğru formda kinetik enerjiye dönüştürmektir.

### *10.1 Sermaye birikimi ≠ üretim artışı*

Klasik ekonomi, sermaye birikimini büyümenin öncülü olarak görür. Oysa Mühendislik Ekonomisi'nde sermaye birikimi, potansiyel enerji deposudur — ancak dönüşüm mekanizması doğru çalışmazsa, bu enerji sistemde âtıl kalır. Türkiye'de geçmişte finansal sermaye birikimi olmasına rağmen üretkenlik artışı düşük kaldı, çünkü motorun "mekanik aktarımı" zayıftı: AR-GE yoktu, risk sermayesi azdı, bilgi dönüşümü sınırlıydı.

### *10.2 Motorun üç ana dişlisi*

Bir ekonomi motorunun üç dişlisi, finansal sermaye, teknolojik kapasite ve güven enerjisidir. Bunlardan biri bozulduğunda motor ses çıkarır ama güç üretmez. Bugün Türkiye'de finansal sermaye mevcut; ancak dişliler arasındaki yağ (güven) yetersiz, dişlilerden biri (teknoloji) küçük kaldı. Mühendislik Ekonomisi yaklaşımı, bu motoru yeniden tasarlamak için aşağıdaki üçlü dengeyi önerir:

$$\text{Enerji Akışı} = (\text{Sermaye} \times \text{Teknoloji Katsayısı} \times \text{Güven Katsayısı})$$

Bu denklem, büyümenin salt para ile değil, sistem mühendisliğiyle sağlanacağını gösterir.

### *10.3 Risk Sermayesi: Yakıtın Ateşleme Noktası*

Otoban döneminde yatırımcılar düşük riskli, öngörülebilir motorları severdi. Off-road ortamında ise motorun dayanıklılığı, riskli yakıtı kontrollü yakabilme becerisidir. Bu nedenle klasik kredi sistemi yerine risk sermayesi ateşleme sistemine geçilmelidir.

### *10.4 Enerji Dönüşüm Verimi*

Her mühendislik sisteminde verim, giriş enerjisinin ne kadarının işe dönüştüğünü gösterir. Ekonomide bu, yatırımın üretime dönüşme oranıdır — yani artımlı sermaye-çıktı oranı (ICOR). Türkiye'nin son dönemlerdeki ICOR değerleri, Kore ya da Çin'in yakalama dönemi ortalamalarına kıyasla belirgin biçimde yüksektir; bu da her birim yatırımın daha az çıktı ürettiği, başka bir ifadeyle motorun verim kaybıyla çalıştığı anlamına gelir. Mühendislik Ekonomisi bakışıyla hedef, bu verimi teknoloji, eğitim, veri ve koordinasyon yoluyla iyileştirmektir. Bu, faiz indirimi kadar etkili bir büyüme aracıdır — ama yapısal olarak kalıcıdır.

### *10.5 Yatırım Motorunun Torku*

Motorun gücü (HP) ne kadar önemliyse, torku da o kadar kritiktir. Ekonomide tork, yatırımların yayılma etkisidir (multiplier effect). Yani bir yatırım, sadece kendini değil, etrafındaki ekosistemi de hareket ettiriyorsa sistem güçlüdür. Teknoloji yatırımlarının avantajı, yüksek tork üretmesidir: örneğin bir yapay zekâ şirketi sadece yazılım geliştirir gibi görünür ama bulut altyapısı, veri güvenliği, çip tasarımı, eğitim sektörünü de tetikler.

### *10.6 Yatırım Motorunun Soğutma Sistemi*

Her güçlü motor gibi, yatırım sistemi de soğutma ister. Bu soğutma, şeffaflık, veri paylaşımı ve güvenilir raporlama ile sağlanır. Aksi halde yüksek ısı (belirsizlik) yatırımcıyı yakar. Bu nedenle Mühendislik Ekonomisi modeli, yatırım ve inovasyon sistemini tıpkı bir motor gibi sensörlerle izleyen, veriye dayalı bir denetim altyapısını önerir.

Mühendislik Ekonomisi yaklaşımı, büyümeyi bir "yakıt artırımı" değil, motor verimi optimizasyonu olarak tanımlar.

## *11. İnsan Sermayesi: Yapay Zekâ Çağında Beyin Gücü Politikası*

Off-road dönemi, sadece sermaye ve teknoloji yatırımıyla değil, insan sermayesi ile de şekillenir. Yapay zekâ, robotik, enerji dönüşümü ve ileri üretim sistemleri gibi alanlarda başarı, yetenekli ve teknik donanımı güçlü iş gücüne bağlıdır. Bu yüzden "beyin gücü politikası" ön plana çıkmalıdır.

### *1. Genç ve Teknik İş Gücü Önemi*

- Türkiye gibi ülkelerde nüfus yapısı genç ve dinamik olabilir; bu büyük bir avantajdır.
- Ama bu genç iş gücünün yeni teknolojilere adapte olabilmesi için eğitim ve sürekli öğrenme programları şarttır.
- Örneğin, üniversitelerde yapay zekâ, veri bilimi, ileri robotik ve enerji teknolojileri gibi alanlara odaklanan bölümler güçlendirilmelidir.

***2. Yetenekleri Çekmek ve Tutmak***

- Küresel rekabette yetenekli mühendis ve araştırmacılar, fırsat bulamadıkları ülkelerden ayrılabilir.
- Türkiye, stratejik teknoloji projeleri ve risk yatırım fonları ile desteklenen AR-GE ortamları oluşturarak, bu yetenekleri hem çeker hem de tutar.
- Örnek: ABD veya Avrupa'dan yetenekli AI araştırmacılarını üniversiteler ve teknoloji merkezleri aracılığıyla Türkiye'ye davet etmek.

***3. Üniversite-Sanayi İş Birliği***

- İnsan sermayesi, sadece bireysel yetenekten ibaret değildir; onu üretim ve AR-GE ile entegre etmek gerekir.
- Yani üniversiteler, laboratuvarlarını ve yeteneklerini teknoloji firmalarının projeleriyle birleştirir.
- Böylece hem yeni ürünler geliştirilir hem de öğrenciler gerçek dünya deneyimi kazanır.

***4. Beyin Gücü ile Ekonomik Döngü***

- İnsan sermayesi, bir kez doğru ekosistemde konumlanırsa, teknoloji yatırımlarının katlanarak büyümesine yol açar.
- Bu döngü, off-road ekonomisinde hız ve adaptasyon yeteneğini artırır: iyi eğitimli ve yetkin iş gücü, yatırımları daha verimli kullanır, riskleri azaltır ve inovasyonu sürekli kılar.

Yapay zekâ çağında, teknoloji yatırımlarının kalbi insan sermayesidir. Beyin gücü politikası ile yetenekleri çekmek, eğitmek ve entegre etmek, Türkiye'nin off-road ekonomisinde hem hızını hem de yönünü korumasını sağlar. Teknoloji fonları sermaye ve AR-GE getirir, ama beyin gücü bu yatırımları sürdürülebilir ve verimli hale getirir.

Ancak insan sermayesinin etkin çalışması, liyakate dayalı bir yetkelendirme mekanizmasına bağlıdır. Mühendislik Ekonomisi'nin motoru ne kadar iyi tasarlanırsa tasarlansın, onu çalıştıracak kadroların teknik ve ahlaki yetkeyle donatılması ön koşuldur. Aksi halde sistem ileri taşınmaz; sadece bazal metabolizma düzeyinde ayakta kalır (Ergen, "Kış Geliyor: Türk Devlet Aklı Erken Anladı, Avrupa Neden Anlamadı?", Ekopolitik, 2026). Yetki dağıtılabilir; yetke inşa edilir — ve bu inşa, liyakati büyüten kurumsal yapılarla mümkündür.

## Tarihsel Döngü: Off-road, Otoban ve Tali Yol

Birinci Dünya Savaşı ve sonrasında imparatorlukların çöküşü, savaşın yarattığı yıkım ve hiper-enflasyon dönemleri, küresel ekonomik sistemi ciddi şekilde kırılganlaştırdı. 1929 Büyük Buhran, dünya ekonomisini adeta off-road'a itti; büyüme durdu, ticaret çöktü ve finansal sistemler büyük bir belirsizlik içine girdi.

### *1950–1970: İlk Otoban Dönemi*

İlk otoban dönemi 1950–1970 arasıdır. Endüstriyel Hız Çağı olarak adlandırabileceğimiz bu dönem, özellikle II. Dünya Savaşı sonrası kurulan Bretton Woods sistemi, Marshall Planı ve seri üretim ekonomisiyle karakterizedir. Üretim hızla büyür, enerji ucuzdur ve ticaret genişler. ABD merkezli küresel kapitalizm, düz ve öngörülebilir bir otoban gibidir; otomotiv, çelik ve petrokimya sektörleri üretim ekonomisinin omurgasını oluşturur.

Türkiye ise bu dönemde ağır sanayiye geçiş ile meşguldü ve kendini küresel otoban için hazırlamaya çalışıyordu. Sanayi ve altyapı yatırımları yapılmakta, enerji dağıtımı ve hidroelektrik projeleri hayata geçiriliyordu. Ancak Türkiye hâlâ dışa kapalı bir ekonomiye sahipti; ihracat ağırlıklı olarak tarım ve tekstil ürünlerinden oluşuyordu. Dünya ekonomisi otoban üzerindeyken, Türkiye daha çok tali yoldaydı: büyüme vardı, fakat yol dar, araç kapasitesi sınırlı ve hız kontrolü şarttı. Türkiye, "otomobil" (sanayi) geliştirmeye çalışıyordu, ancak yol ve altyapı hâlâ otoban hızını kaldıracak düzeyde değildi.

### *1970–1980: Tali Yol ve Şoklar Dönemi*

1970'lerde otoban bozulur ve dünya ekonomisi tali yola girer; bu dönem Krizler ve Şoklar Dönemi olarak adlandırılır. Petrol krizleri (1973 ve 1979) üretim maliyetlerini patlatır, Bretton Woods sistemi çöker ve dolar altından ayrılır. Enflasyon ve durgunluk aynı anda görülür; yani stagflasyon yaşanır. Her ülke yeniden denge arayışına girer. Türkiye dahil birçok ülke için bu dönem, dışa kapalı ve planlı ekonomilerin tıkanma dönemidir.

1970–1980 Türkiye'si, petrol şokları ve döviz krizleriyle sarsılmış bir tali yol ekonomisiydi. Otoban hızından uzak, yol bozuk ama hâlâ ilerlemeye çalışan bir ekonomik yapı mevcuttu. Ekonomik kırılganlık ve dış şoklar, siyasi çalkantı ve toplumsal gerilim ile birleşti. Koalisyon hükümetleri, grevler ve siyasi şiddet, reform ve istikrar çabalarını sekteye uğrattı.

***1980–2008: Yeni Otoban — Neoliberal Dönüşüm ve Küreselleşme***

1980–2008 dönemi, Yeni Otoban olarak adlandırılabilir. Reagan ve Thatcher dönemiyle başlayan neoliberal dönüşüm, ekonomiyi yeniden açtı ve piyasa mekanizmalarını güçlendirdi. Mikroçip devrimi, kişisel bilgisayar ve iletişim teknolojileriyle birlikte yeni bir üretim yapısı ortaya çıktı. Bu dönemde hızlı hareket eden ülkeler ve şirketler, yeni ekonomiden önemli paylar elde etti. Çin, Japonya ve Malezya çip üretiminde; Hindistan ise yazılım sektöründe küresel üretim zincirine dahil oldu.

1990'lara gelindiğinde internet ekonomisi, bu otobanı genişleterek adeta bir tüketici elektroniği otobanına dönüştürdü. Sermaye hareketleri serbestleşti; Çin, Hindistan ve Doğu Avrupa ülkeleri üretim zincirine aktif olarak katıldı. Her evde televizyon, bilgisayar, cep telefonu ve kredi kartı yaygınlaştı.

Türkiye için asıl sıçrama 2001 krizinden sonra geldi. Ciddi bir makroekonomik istikrar programı uygulandı: enflasyon düşürüldü (%70'lerden %30'lara geriledi, 2005 civarında tek haneli seviyelere yaklaştı). Merkez Bankası bağımsızlaştı, faiz politikaları disipline girdi. Kamu maliyesi reformlarıyla bütçe açıkları kontrol altına alındı. 2002–2008 dönemi boyunca ortalama büyüme %5–7 civarında gerçekleşti. Bankacılık sektörü reformu sağlamlaştırıldı, özelleştirmeler hızlandı, AB uyum süreci reformları hızlandırdı, yabancı yatırımcılar Türkiye'ye yöneldi.

Yani dünya 1980'den itibaren Yeni Otoban'daydı ama Türkiye bu otobana ancak 2000'lerde çıkabildi. 1990'ların kırılgan tali yolundan, yapısal reformlar ve siyasi istikrar sayesinde 2000'lerde otobana geçiş yapıldı.

***2008–2019: Tali Yol — Kırılgan Büyüme Dönemi***

2008 küresel finansal kriziyle birlikte Yeni Otoban dönemi sona erdi ve dünya ekonomisi tali yola girmeye başladı. Kriz, küresel finans sisteminin dengesizliğini açıkça ortaya koydu. Gelişmiş ekonomiler, düşük faiz ve parasal genişlemeyle geçici bir "yama" uyguladı. Ancak büyüme, altyapı yatırımı yerine likiditeyle desteklenen bir büyümeydi. Ekonomi hâlâ ilerliyordu, fakat artık tali yoldaydı: hız düşük, yol bozuk ve araçlar hassastı.

Bu dönemin tipik özellikleri Çin'in devasa ihracat gücü, ABD'nin borçlanarak talep yaratması ve Avrupa'nın durgunlaşmasıdır.

***2020 Sonrası: Off-road — Bilinmeyen Arazi***

Covid-19 salgını, 2008'den bu yana süren tali yolu kapatıp ekonomiyi tam anlamıyla off-road'a soktu. Dijital dönüşüm, yapay zekâ ve enerji geçişi, ekonomiyi bilinmeyen bir arazide ilerletmeye zorladı. Ekonomi artık düz bir yol üzerinde değil; engebeli, riskli ve yönsüz bir coğrafyada hareket ediyor. Bu dönüşümün ölçeğini yukarıda 2020 bölümünde ayrıntılı olarak ele aldık.

Özetlersek dört dönem: 1950–1970 (İlk Otoban), 1970–1980 (Tali Yol / Şoklar), 1980–2008 (Yeni Otoban), 2008–2019 (Tali Yol / Kırılgan Büyüme) ve 2020 sonrası (Off-road). Türkiye'nin küresel ritme tam uyumu yalnızca 2002–2008 penceresinde gerçekleşti.

## Orta Gelir Tuzağı ve Güney Kore Örneği

Orta gelir tuzağından çıkmak için örnek gösterilen ülkelerden Güney Kore başta geliyor. Güney Kore, düşük gelirli bir ekonomiden yüksek gelirli bir ekonomiye dönüşen birkaç ülkeden biri. OECD'nin Kalkınma Yardım Komitesi'nden (DAC) yardım alan bir ülkeden, DAC'a yardım eden bir ülkeye dönüşen tek ülkedir.

Güney Kore ile aynı nüfus oranına sahip ve aynı koşullarda olan Türkiye de Güney Kore gibi orta gelir tuzağından çıkabilir deniliyor. Neredeyse aşağı yukarı aynı benzerlikleri gösteren bir süreci yaşamışız — tek farkımız onların Japon kolonileşmesine maruz kalmaları, bizim herhangi bir başka yönetim altında kalmamamız ve geçmişten gelen büyük bir imparatorluğun bakiyesini taşımamız. Her ne kadar kolonileşmeyle kazanımlar elde edildiği söylense de ekonomistlere göre bu süreçte yalnızca ilköğretime yatırım yapılmış (çocukların sadece %47'si okula kayıtlıymış) ve yapılan altyapı yatırımları Kore savaşı sırasında yerle bir olmuş. Aynı şekilde bizim ülkemiz gibi yeraltı kaynaklarına sahip değiller; bu kaynaklar Kuzey Kore'de kalmış.

Bugün ülkemizin orta gelir tuzağından çıkması, ekonominin düzelmesi için yapısal reformların yapılması gerektiği söyleniyor. Hangi yapısal reformlar? Sadece yargı bağımsızlığı, hukukun üstünlüğü, kurulların şeffaflığı, ihale kanunu gibi yapısal reformlar yeterli mi, ya da gerekli mi? Yabancı yatırımcının gelmesi beraberinde kalkınmayı da getirir mi, orta gelir tuzağından çıkarır mı, yoksa 2010'lu yıllara kadar süreçte gördüğümüz gibi sadece parametreleri mi iyileştirir?

Kalkınmada Amerika ile başlanırsa sonucu Latin Amerika ülkelerinde görebiliriz. Washington Konsensüsü onlar için bir reçeteydi. Washington Konsensüsü; mali disiplin, özel mülkiyetin korunması, kamu harcamalarının azaltılması, kamu teşebbüslerinin özelleştirilmesi, vergi reformu, ticaretin serbestleştirilmesi, finansal reform, uluslararası ticaretin önündeki engellerin kaldırılması, sermaye hareketlerinin liberalleştirilmesi, serbest faiz hadleri, rekabetçi kur politikaları ve güçlü fikri haklar rejimi olarak tanımlandı.

Burada iki akademik tezin birlikte değerlendirilmesi gerekir. Rodrik, Stiglitz ve Acemoğlu'na göre Washington Konsensüsü başarısız oldu çünkü kurumsal zemin zayıftı. Keun Lee'ye göre ise kurumlar tek başına yetmez; asıl eksik teknoloji politikaları ve yükseköğrenim reformuydu. Bu iki tez birbirini dışlamaz, tamamlar: sağlam kurumlar gereklidir ama yeterli değildir; Lee'nin eklediği teknoloji politikası eksik halkadır. Bu nedenle orta gelir tuzağından çıkış için hem yargı, şeffaflık, ihale kanunu gibi kurumsal reformlar şarttır hem de bunun üzerine inşa edilecek bir ulusal inovasyon sistemi.

Rodrik'e göre Washington Konsensüsünün önerilerini Latin Amerika ülkeleri aynı anda uyguladı. Doğu Asya ülkeleri ise zamana yayarak uyguladı. Buna ek olarak, Doğu Asya ülkeleri Latin Amerika'dan farklı bir Ulusal İnovasyon Sistemi ile Bilim ve Teknoloji Politikası uyguladı. Bugün Latin Amerika, çeşitli sosyoekonomik ve güvenlik sorunlarıyla mücadele eder hale geldi; ekonomik sıkıntılar, yüksek enflasyon, artan suç oranları, yolsuzluk, eğitim yetersizliği, plansız şehirleşme ve düşük yatırımcı ilgisi bu tablonun ana çizgileridir.

***Güney Kore Örneği***

Güney Koreli ekonomist Prof. Keun Lee'ye göre son birkaç on yılda, Güney Kore, otoriter bir rejimden ve yarı kapalı, devlet odaklı bir ekonomiden güçlü bir demokrasiye ve çok açık bir pazar ekonomisine dönüşmeyi başardı. Güney Kore bu dönüşümü, Washington Konsensüsünün reçetesini körü körüne uygulayarak değil, gerektiğinde yavaş ilerleyerek ve hesaplanmış sapmalar / yan yollar yaratarak gerçekleştirdi. Lee'ye göre Washington Konsensüsünün başarısız sonuçları yalnızca kurumların zayıflığı değil, teknoloji politikalarının olmayışı ve yükseköğrenim reformunun olmayışıdır.

1950'lerde ve 1960'larda, Güney Kore gıda yardımı alarak hayatta kalırken, sanayileşmeden önce halkın beslenmesini önceliklendirmiş. 1961'de askeri hükümet, hububat için çiftçilerden yüksek fiyatla alıp tüketicilere düşük fiyatla satarak çift fiyat politikası uygulamış. Bu süreçte, yüksek getirili pirinç türleri geliştirilmiş ve bankalar millileştirilmiş. Mali ve ticari liberalizasyona temkinli yaklaşarak, sanayileşmeyi destekleyecek şekilde iç tasarrufları artırmaya odaklanmışlar.

Güney Kore'de bankalar, hükümetin yaklaşık yirmi yıl boyunca düşük faiz politikasıyla yatırımları teşvik etmesi ve tasarrufların imalat sektörünün kapasitesini artırmaya yönlendirilmesini sağlamasından sonra özelleştirilmiş. Bu süreç, gelirlerde o kadar güçlü bir artışa katkıda bulunmuş ki, baskılanan faiz oranlarına rağmen, iç tasarruf oranı 1960'ların başında GSYİH'nın yaklaşık %3'ünden 1980'lerin sonunda yaklaşık %36'ya yükselmiş (Bank of Korea).

Güney Kore, ticaretin serbestleştirilmesine de benzer şekilde ihtiyatlı bir yaklaşım benimsemiş. Bir ülke ticareti serbestleştirdiğinde, yerel firmaların yabancı şirketlerle rekabet edebilmesi gerekir. Aksi takdirde, yabancı şirketler eğer hükümet koruyucu önlemler almazsa tekel kurabilir veya yerel sanayi altyapısını yok edebilir. Güney Kore'de bu koruma, tüketim mallarına uygulanan çok yüksek gümrük vergileri şeklinde gelmiş. Korunan yerel firmalar, tekel gelirlerini yatırımları finanse etmek için kullanmışlar, çünkü hükümetin koruma önlemleri ihracat performansına bağlıymış ve firmalar yine de dünya ihracat pazarlarının disiplinine tabi tutulmuş. Bu arada, Güney Kore'nin ithal etmek zorunda olduğu sermaye mallarına ise çok düşük gümrük vergileri uygulanmış.

Güney Kore'nin başarı hikayesi, otoriter bir rejim altında insan sermayesine yapılan yatırımlar ve ihracata yönelik büyüme stratejilerinin eş zamanlı olarak uygulanması ile şekillenmiştir. Ülke, ithalatı ikame etme politikalarını da uygulamış ve yabancı rekabetten korunmak için yüksek tarifeler koymuş. Tüm bu stratejiler, Güney Kore'nin gelişiminde en etkili yolu izleyerek, demokratik ve açık bir ekonomi olmasına katkı sağlamıştır.

Önce demokrasi sonra gelişmişlik mi, önce gelişmişlik beraberinde demokrasi mi? Güney Kore nasıl yakaladı? Yakalama, Schumpeter ekolüne göre geriden gelen ülkelerin gelişmiş ülkeler ile arasındaki farkları kapatması olarak ya da gelişmiş ülkeleri imitasyon ile takip etmesi olarak tarif edilmiştir. Lee'ye göre gelişmiş ülkeler gibi olmak istiyorsan farklı ol. Uzun vadeli başarı, gelişmiş ülkelerin izlediği yoldan farklı bir yol izlemekle olur. Bu Zeno Paradoksuna benzetilir: bir kaplumbağayı yakalamak istiyorsan onun olduğu yeri hedeflersen hiçbir zaman yakalayamazsın, çünkü kaplumbağa ileride bir noktaya gitmiş olur.

## Ülkemizde AR-GE İştahı

Nasıl yakalayacağız? Biz ve diğer orta gelir tuzağındaki ülkelerde, özel sektörün genellikle holdinglerin ve KOBİ'lerin elinde olması belirgin bir özellik. Holdingler ister istemez ihtiyaçtan zamanla ortaya çıkmış yapılar. Bu yapıların en önemli özelliği iç pazarın paylaşımında etkin rol oynaması.

Burada önemli bir nokta var: Holding yapısının kendisi sorun değildir. Kore'nin çaebol'leri de benzer çok-iş-kollu yapılardı ve Kore kalkınmasının motoru oldular. Türk holdinglerini sorunlu kılan yapı değil, yönlendirilme biçimidir. Çaebol'ler ihracata ve kendi AR-GE'lerini üretmeye zorlandı; Türk holdingleri ise AR-GE ihtiyacını yabancı ortaklarından karşıladı, dolayısıyla AR-GE iştahı neredeyse oluşmadı ve genellikle göstermelik kaldı. Birden fazla iş koluna dağılmaları da yerleşik olma ihtiyaçlarındandır.

Yeni bir yabancı şirket iç piyasaya girerken bu holdinglerle iş birliği yaparak girer ve holdingin diğer iş kolundaki şirketleri buna kaldıraç etkisi yaratır ya da riski yayar. Genellikle iktidar ile yakın çalışma ihtiyacı içindedirler. KOBİ'ler ise düşük katma değerli üretirler ve sadece istihdam yaratan kuruluşlar olarak görülür. Örneğin ülkemizin %99,7'si KOBİ'dir ve istihdamın %70,5'ini sağlar (TÜİK, 2023).

Kalkınmanın en önemli diğer aktörü üniversitelere gelince. Üniversiteler ile endüstrinin kopukluğunu bu ülkelerde görebiliriz. AR-GE'ye ihtiyacı olmayan bir endüstri, ister istemez üniversite-sanayi iş birliğini kadük bırakır. KOBİ'lerin düşük katma değerli, popüler olmayan AR-GE'ye olan ihtiyacı da üniversitenin iştahını karşılamaz.

Bu üniversitenin kendi içine kapanmasına ve temel veya popüler konulara yönelmesini beraberinde getirir. Doktorasını yurt dışından almış veya gelişmiş ülkelerin popüler konularını takip eden akademisyenler ister istemez bu konular etrafında çalışmaya devam ederler ve ülkenin yarattığı bilim çıktısı ulusal endüstrinin bir katkısı olmaz. Verilen derslerden yapılan araştırmalara kadar üniversite çıktısı sanayiyi teğet geçer. Katma değer yaratamayan üniversite ise devlet ve endüstri tarafından bir istihdam yetiştirme yeri olarak görülür. AR-GE kaynakları gittikçe azalır, öğretim üyelerinin ders yükü ve öğrenci sayısı da artar. Üniversiteler, yüksek okullara yakınsamaya başlar.

Orta gelir tuzağındaki ülkelerin temel çıkmazı buradadır. Yüksek katma değerli ürün veya şirket yaratamazlar. Bugün bunların yansımaları AR-GE'ye yatırılan miktarın azalmasıyla, düşük katma değerli sektörlerin endüstriyi ele geçirmesiyle, nitelikli iş gücünün olmamasıyla, beyin göçünün başlamasıyla, şeffaflığın kalkmasıyla, bürokratik engellerin artmasıyla, döviz kurunun dalgalanmalarıyla önümüzde duruyor.

Şimdi bu konularda iyileştirme yapılsa da üniversite-sanayi iş birliği çıkmazı kırılmadıkça ülkenin orta gelir tuzağından çıkma ihtimali gözükmüyor. AR-GE teşviklerini artırsanız bu endüstri yapısı kullanamaz; araştırma bütçelerini artırsanız çıktılar gelişmiş ülkelerin konularına gider; araştırmalarla katma değeri yüksek teknoloji bulsanız holding yapısı ürünleştiremez, KOBİ'nin gücü yetmez; kurumsal yapıları şeffaflaştırsanız yurt dışı monopollerinin önü açılır, yerli ekonomi silinir; Güney Kore'deki gibi düşük faize dayansanız bunu kullanıp ihracat yapacak KOBİ ve holding bulamazsınız çünkü hepsi ithalata dayalı ihracat içinde olduğu için kur patlar.

Bu tespitler ışığında eğer yüksek teknoloji temelli bir kalkınma hedeflenecek ise Lee'nin tanımladığı iki eksik bir bariyer ile karşılaşılır ve buna değinmek elzemdir. İlk eksik, AR-GE ihtiyacı olan büyük şirketin olmayışıdır. Böyle büyük şirket olsa üniversiteleri çalıştırır, startupları satın alır ve girişimcilik çarkını ilerletir. İkinci eksik ise odaklanılacak teknoloji konusunda yetişmiş insan kitlesinin hazır olmamasıdır. Bariyer ise herkesten sonra girdiğiniz konularda gelişmiş ülkelerin uyguladığı fikri haklar rejimidir.

Bunun için Lee'ye göre yan yol seçilmeli, kısa döngülü teknolojiler üzerine fırsat penceresi kollanmalıdır. AR-GE teşvikleri proje temelli değil, yetenek yetiştirmek üzerine kurulmalı; AR-GE'ye aç büyük şirketlerin çıkması için destek sağlanmalı ya da bu tür şirketler satın alınmalıdır. 3–5 yıllık denemeler ile bu döngü başarı olana kadar tekrarlanmalıdır. Bir teknolojinin döngüsü ilk patent ile ona atıf yapan ilk patent arasındaki zamandır. Uzun döngülü teknolojiler gelişmiş ülkelerin konuları olarak kalmalı, popüler konulardan öte fikri hakları oluşmamış kısa döngülü konular seçilmelidir.

Bu makro plan, Amerika'nın dünyaya verdiği girişimci ekonomisi ile mikro ölçekte desteklenmesi gerekir. Bu ekonominin aktörleri risk sermayesi, risk yatırımcısı ve girişimcidir. Bu üçlü, büyük şirketleri çıkarmak, büyük şirketlerin ihtiyacı olan startupları çıkarmak, üniversitelerin bilimsel çıktılarını inovasyona dönüştürmek ve üniversitelerin yetiştireceği insan kaynağını kullanmak için ihtiyaç olan döngüyü kuracaktır. Aynı zamanda gelişmiş ülkeler ile iş birliği ve gelişmekte olan ülkelere de açılma ile sinerji yaratacaktır.

## Teorik Çerçeve: Büyüme İktisadının Nobel Ödüllü Öncüleri ile Karşılaştırma

Bu makalenin önerdiği kalkınma çerçevesi, modern büyüme iktisadının Nobel ödüllü öncülerinin teorik temelleri üzerine inşa edilmiştir. Aşağıda, makaledeki ana tezlerin bu kuramcıların çalışmalarıyla nasıl örtüştüğü ve nerede farklılaştığı değerlendirilmektedir.

Kurumların Belirleyici Rolü — Daron Acemoğlu ve James Robinson (Nobel 2024): Makalenin Washington Konsensüsü eleştirisi ve kurumsal zeminin zayıflığına yaptığı vurgu, Acemoğlu ve Robinson'un "kapsayıcı kurumlar" (inclusive institutions) olmadan sürdürülebilir büyümenin mümkün olmadığı tezine doğrudan yaslanmaktadır. Ancak makale bir adım öteye giderek, kurumsal reformların gerekli ama yeterli olmadığını savunur. Acemoğlu'nun çerçevesine Keun Lee'nin teknoloji politikası ve ulusal inovasyon sistemi eksikliğini ekleyerek tamamlayıcı bir eleştiri sunar: sağlam kurumlar şarttır, ancak üzerine inşa edilecek bir inovasyon ekosistemi olmadan kalkınma tamamlanamaz.

İçsel Büyüme ve AR-GE'nin Motoru — Paul Romer (Nobel 2018): Makalenin merkezindeki "büyümenin motoru finansman değil, teknoloji-yetenek-risk sermayesi üçgenidir" tezi, Romer'in içsel büyüme teorisinin doğrudan bir uygulamasıdır. Romer, büyümenin kaynağının fiziksel sermaye birikimi değil fikirler ve AR-GE olduğunu kanıtlamıştır. Makale, Romer'in modeline gelişmekte olan ülkelere özgü pratik kısıtları ekler: fikri haklar bariyeri, kısa döngülü teknoloji stratejisi ve AR-GE teşviklerinin proje temelli değil yetenek yetiştirme odaklı olması gerektiği önerisi, Romer'in teorisini operasyonel bir politika çerçevesine dönüştürür.

Beşeri Sermaye ve Beyin Göçü — Robert Lucas Jr. (Nobel 1995): Lucas, beşeri sermaye birikiminin ülkeler arası gelir farklılıklarını açıkladığını ve "neden sermaye zengin ülkelerden fakir ülkelere akmıyor?" sorusunu gündeme getirmiştir. Makaledeki üniversite-sanayi kopukluğu analizi, beyin göçü sorunu ve nitelikli iş gücünün yokluğu, Lucas'ın beşeri sermaye çerçevesinin Türkiye özelindeki doğrulamasıdır. Makalenin eğitim reformu ve "hata yapmaktan korkmayan bireyler yetiştirme" önerisi, Lucas'ın temel tezini politika düzeyinde somutlaştırır.

Küreselleşme Eleştirisi ve Devletin Stratejik Rolü — Joseph Stiglitz (Nobel 2001): Stiglitz, Washington Konsensüsünü en sert eleştiren Nobel ödüllü ekonomisttir. Makaledeki "erken liberalizasyon gelişmekte olan ekonomileri savunmasız bırakır" argümanı ve Güney Kore'nin koruyucu politikalarla sanayileşme hikâyesi, Stiglitz'in "Globalization and Its Discontents" tezleriyle güçlü bir paralellik taşır. Stiglitz, bilgi asimetrisi ve piyasa başarısızlıkları nedeniyle devletin düzenleyici rolünün vazgeçilmez olduğunu savunurken, makale bu yaklaşımı "Mühendislik Ekonomisi" kavramıyla operasyonel bir çerçeveye taşır: regülasyonlar sabit kurallar değil, zeminin koşullarına göre ayarlanan adaptif mekanizmalardır.

Ölçek Ekonomileri ve Stratejik Ticaret — Paul Krugman (Nobel 2008): Krugman'ın "yeni ticaret teorisi", ölçek ekonomilerinin ve ağ etkilerinin ticaret kalıplarını belirlediğini göstermiştir. Makalenin "ülke sınırlarını aşabilecek, monopol vizyonu olan büyük teknoloji şirketleri çıkarmak" hedefi, Krugman'ın ölçek ekonomileri argümanının doğrudan yansımasıdır. Holding yapısının iç pazara odaklı kalması ve küresel ölçekte rekabet edebilecek firmalar üretememesi sorunu, Krugman'ın çerçevesinde stratejik ticaret politikasının neden gerekli olduğunu açıklar.

Yapısal Dönüşüm ve İkili Ekonomi — W. Arthur Lewis (Nobel 1979): Lewis'in ikili ekonomi modeli, geleneksel düşük verimlilikli sektörden modern yüksek verimlilikli sektöre iş gücü transferini kalkınmanın temel mekanizması olarak tanımlar. Makaledeki holding-KOBİ-üniversite üçgeninin kopukluğu ve düşük katma değerli üretimden yüksek katma değerli teknoloji firmalarına geçiş çağrısı, Lewis'in yapısal dönüşüm kavramının çağdaş bir versiyonudur. KOBİ'lerin istihdamın büyük bölümünü sağlamasına rağmen düşük katma değerli kalması, Lewis'in "sınırsız emek arzı" aşamasının Türkiye'de hâlâ aşılamadığının göstergesidir.

Yaratıcı Yıkım ve Girişimci Ruh — Joseph Schumpeter: Her ne kadar Nobel ödülünü alamadan hayatını kaybetmiş olsa da Schumpeter, büyüme iktisadının kurucu figürlerinden biridir. Makalenin risk sermayesi-girişimci-inovasyon üçlüsü ve "hata yapmaktan korkmayan, subjektif ama vizyoner kararlar" çağrısı, Schumpeter'in "yaratıcı yıkım" (creative destruction) ve "Unternehmergeist" (girişimci ruhu) kavramlarının modern bir yorumudur. Makale aynı zamanda Schumpeter ekolünün "yakalama" (catching up) kavramına Lee üzerinden doğrudan atıf yapar: gelişmiş ülkeleri yakalamak istiyorsan onların izlediği yoldan farklı bir yol izlemelisin.

Makalenin Özgün Katkısı — Mühendislik Ekonomisi Paradigması: Yukarıdaki kuramcıların her biri, kalkınmanın farklı bir boyutunu aydınlatır: Acemoğlu kurumları, Romer fikirleri, Lucas beşerî sermayeyi, Stiglitz devletin rolünü, Krugman ölçeği, Lewis yapısal dönüşümü, Schumpeter ise girişimciyi merkeze alır. Bu makalenin özgün katkısı, tüm bu unsurları bir araya getiren "Mühendislik Ekonomisi" paradigmasıdır. Ekonomi yönetimini bir kontrol mühendisliği problemi olarak kavramsallaştıran bu yaklaşım, regülasyonları sabit kurallar yerine akıllı süspansiyon sistemi gibi ele alır: zemine göre ayarlanır, veriye dayanır ve sistemi yolda tutmayı amaçlar. Liberal ekonominin minimum devlet müdahalesinden, karma ekonominin piyasa-kamu dengesinden farklı olarak Mühendislik Ekonomisi, günümüzün belirsiz küresel ortamında ekonomiyi statik bir denge problemi değil, sürekli ayar gerektiren dinamik bir sistem olarak ele alır. Makro ölçekte Kore'nin sanayi disiplinini, mikro ölçekte Silikon Vadisi'nin girişimci ekosistemini hedefleyen bu sentez, yukarıdaki Nobel ödüllü kuramcıların teorik mirasını Türkiye'ye özgü bir operasyonel çerçeveye dönüştürmektedir.

## Sonuç

Türkiye'nin orta gelir tuzağından çıkışı, yalnızca daha fazla üretmekle değil, nasıl ve ne için ürettiğini yeniden tasarlamakla mümkündür. Bu dönüşümün merkezinde risk sermayesinin oluşumu, girişimci ekosistemin sürekliliği ve ekonominin statik değil dinamik bir mühendislik sistemi olarak ele alınması yer almalıdır.

Öncelikle, eski sermayenin kazancının girişimci sermayeye yönelmesini teşvik eden bir mekanizma kurulmalıdır. Risk yatırım ekosistemi sabır ister; bol hata, uzun vadeli bakış ve subjektif ama vizyoner karar alma becerisi gerektirir. Bu nedenle girişimciliği yalnızca bir sonuç değil, ilköğretimden itibaren desteklenen bir döngü olarak ele almak zorundayız. Eğitim sistemi; hata yapmaktan korkmayan, deneme-yanılmayı öğrenmenin parçası sayan bireyler yetiştirmeli, bu zihniyet finansman ve regülasyonla tamamlanmalıdır.

Bu noktada odaklı teşvikler ve subjektif karar verebilen kurumsal yapılar kritik önemdedir. Potansiyel vadeden şirketlerin vergisel yüklerden geçici olarak muaf tutulması, özsermaye yapılarının güçlendirilmesi ve yatırımlarla sistematik biçimde desteklenmesi gerekir. Amaç, küçük ve kırılgan şirketler değil; ülke sınırlarını aşabilecek, monopol vizyonu olan büyük teknoloji şirketleri çıkarmaktır. Nitekim insansız uçaklar, dijital oyun ve e-ticaret gibi alanlarda bu pırıltıların emareleri şimdiden görülmektedir. Bu tür şirketler, tereddütsüz ve sonuna kadar desteklenmelidir.

Bu ivmenin sürdürülebilir olması için üniversiteler de dönüşmelidir. Üniversiteler özerkleştirilmeli, müfredat ve araştırma alanlarında serbest bırakılmalıdır. Devletin rolü doğrudan müdahale değil; AR-GE finansmanı, performans bazlı fonlama ve öğrencilere sağlanan katkı payları yoluyla dolaylı ama güçlü bir yönlendirme olmalıdır.

Tüm bu çabanın üzerinde ise daha kapsayıcı bir paradigma yer almalıdır: Mühendislik Ekonomisi. Liberal ekonomi minimum devlet müdahalesini, karma ekonomi ise piyasa ile kamunun dengesini savunur. Mühendislik Ekonomisi ise bunlardan farklı olarak, günümüzün belirsiz ve riskli off-road küresel ortamında ekonomiyi statik bir denge problemi değil, sürekli ayar gerektiren dinamik bir sistem olarak ele alır. Bu yaklaşımda regülasyonlar birer sabit kural değil, akıllı süspansiyon sistemi gibidir: zemine göre ayarlanır, veriye dayanır ve sistemi yolda tutmayı amaçlar.

- Enflasyon ve para politikası, Hazine ve Merkez Bankası arasında model tabanlı, hızlı adapte olabilen planlarla koordine edilir.
- Bütçe açıkları ve borçlanma, bir zafiyet değil; getirisi ölçülebilir, proje bazlı yatırımlar için kullanılan stratejik mühendislik araçlarına dönüşür.

- Döviz kuru riskleri, proaktif araçlar, stres testleri ve simülasyonlarla yönetilir; ithalat-ihracat yapısı ve likidite dengesi sürekli optimize edilir.
- Uzun vadeli stratejik yatırımlar (AR-GE, enerji, savunma, dijital dönüşüm), risk yatırım fonları ve teknoloji odaklı sermaye ile entegre edilir. Holdingler ana işlerine odaklanırken, sermaye esnek biçimde yeni teknolojik alanlara akar.

Sonuç olarak, 2000'lerin "otoban" ekonomisi sona ermiştir. Bugünün dünyası bozuk, kaygan ve öngörülemez bir off-road parkurudur. Bu zeminde başarı; yalnızca hızdan veya güçten değil, doğru araç, doğru tasarım ve hızlı reflekslerden geçer. Makro ölçekte Kore'nin sanayi disiplinini, mikro ölçekte Silikon Vadisi'nin girişimci ekonomisini hedeflemeli; hata yapmaktan korkmayan, subjektif ama vizyoner kararlar alabilen bu yapıyı başarana kadar birlikte işletmeliyiz. Türkiye'nin avantajı; rakiplerinden daha büyük olmak değil, değişen koşullara rakiplerinden daha hızlı adapte olabilme kapasitesini kurabilmesidir. Bu, tam da "kurt refleksi"dir: kışı durdurmaya çalışmak yerine kışa göre değişmek, tehlikeyi erken sezip gürültü yapmadan hazırlanmak ve enerjiyi doğru zamanda kullanmak. Ancak bu refleksin ölçeklenmesi, güçlü liderliğin teknik ve ahlaki yetkeye sahip kadrolarla buluşmasıyla mümkündür (Ergen, "Kış Geliyor", Ekopolitik, 2026).

## Kaynakça

*Not: Claude, ChatGPT, Gemini, Grok gibi yapay zekâ araçlarından yardım alınmıştır.*